\documentclass[]{pasj02} 
\usepackage[switch,mathlines]{lineno} 
\usepackage{natbib} 
\usepackage[T1]{fontenc}
\usepackage{amsmath}
\usepackage{multirow}
\usepackage{xcolor}

\jyear{2026}
\Received{}
\Accepted{}

\newcommand{\Sigmag}{\Sigma_\mathrm{g}}
\newcommand{\Sigmad}{\Sigma_\mathrm{d}}
\newcommand{\del}{\partial}
\newcommand{\vgr}{v_{{\rm g},r}}
\newcommand{\vdr}{v_{{\rm d},r}}
\newcommand{\St}{{\rm St}}
\newcommand{\cs}{c_{\rm s}}
\newcommand{\vK}{v_{\rm K}}
\newcommand{\pmid}{p_{\rm mid}}
\newcommand{\Tmid}{T_{\rm mid}}
\newcommand{\nut}{\nu_{\rm t}}

\begin{document}

\title{Thermal instability and rocky planetesimal formation in the inner regions of protoplanetary disks}

\author{
Ryo \textsc{Kato}\altaffilmark{1}\altemailmark\orcid{0009-0007-6867-1459}\email{kato.r.be32@m.isct.ac.jp},
Takahiro \textsc{Ueda}\altaffilmark{2}\orcid{0000-0003-4902-222X},
Satoshi \textsc{Okuzumi}\altaffilmark{1}\orcid{0000-0002-1886-0880}
}

\altaffiltext{1}{Department of Earth and Planetary Sciences, Institute of Science Tokyo, Meguro, Tokyo 152-8551, Japan}
\altaffiltext{2}{National Astronomical Observatory of Japan, 2-21-1 Osawa, Mitaka, Tokyo 181-8588, Japan}



\maketitle

\begin{abstract}
The inner regions of protoplanetary disks are promising formation sites of rocky planetesimals. Theoretical studies have proposed that dust trapping at the magnetorotational instability (MRI) activation boundary, or the dead-zone inner edge, promotes planetesimal formation. However, the inner disk may be thermally unstable, in which case the dead-zone inner edge may not remain steady, and the associated pressure maximum and dust trap may not be maintained. In this study, we propose a scenario in which planetesimals form in a thermally unstable inner disk through dust self-accumulation driven by the coevolution of dust and disk temperature. To this end, we simultaneously calculate the non-equilibrium thermal evolution, the evolution of the gas and dust surface densities, dust growth, and planetesimal formation. Our results show that thermal instability triggers cyclic MRI activation and deactivation, during which planetesimals form. The MRI is activated in the inner disk, and thermal instability causes the active region to expand outward and then return to an inactive state, producing a periodic cycle. Triggered by a local enhancement in the dust surface density, dust undergoes self-accumulation while migrating inward during the MRI-inactive phase, resulting in planetesimal formation. Once the MRI is reactivated at a smaller radius, the next cycle begins. For a typical accretion rate of $10^{-8}M_{\odot}~{\rm yr^{-1}}$, a planetesimal belt forms near 1 au. Depending on the model parameters, approximately 10--80\% of the dust flowing into the planetesimal-forming region is converted into planetesimals. This mechanism produces sufficient planetesimal mass for the formation of multiple super-Earths. The resulting planetesimal distribution can serve as a physically motivated initial condition for subsequent planet formation simulations.
\end{abstract}


\section{Introduction}\label{sec:intro}
The inner regions of protoplanetary disks are the sites of rocky planet formation. 
In the solar system, the four terrestrial planets are distributed within 0.39--1.5~au.
Exoplanet observations suggest that close-in super-Earths are common, and that a subset of them have bulk densities consistent with rocky compositions \citep[e.g.,][]{Dressing&Charbonneau2015, Weiss&Marcy2014, Marcy+2014}.
To understand the origin of these planetary systems, it is essential to clarify how planetesimals form in the inner regions of protoplanetary disks.

During the dust growth stage, sticking growth is inhibited by rapid radial drift caused by the negative pressure gradient along the radial direction \citep{Whipple1972, Adachi+1976, Weidenschilling1977}. 
A local maximum in the gas pressure profile can halt and reverse dust drift, thereby producing efficient dust trapping \citep{Brauer+2008, Pinilla+2012, Drazkowska+2013, Ueda+2019, Ueda+2021}.
If dust accumulation increases the midplane dust-to-gas mass ratio to order unity, dust settling can be further enhanced and may cause planetesimal formation via gravitational instability \citep[e.g.,][]{Goldreich&Ward1973, Sekiya1998, Youdin&Shu2002}.
The streaming instability, driven by aerodynamic dust–gas feedback \citep{Youdin&Goodman2005, Johansen&Youdin2007}, is also a promising mechanism for planetesimal formation. However, it remains unclear whether the streaming instability can develop sufficiently at pressure maxima to initiate planetesimal formation.

One plausible site for rocky planetesimal formation is the inner edge of the dead zone \citep{Gammie1996}, where the gas temperature reaches $\sim 1000~\rm{K}$, above which thermal ionization of alkali metals activates the magnetorotational instability (MRI) \citep{Desch+2015}.
The pressure maximum at the dead-zone inner edge has been proposed as a site of solid accumulation and planet formation \citep[e.g.,][]{Kretke+2009,Dzyurkevich+2010, Flock+2017}. 
Early two-dimensional simulations also demonstrated rapid protoplanetary embryo formation through solid accumulation at the dead-zone inner edge \citep{Lyra+2008,Lyra+2009}.
An application of this pressure trap to compact planetary systems is the inside-out planet formation scenario of \citet{Chatterjee&Tan2014}, in which inward-drifting pebbles are trapped at the dead-zone inner edge and form close-in planets sequentially from the inside out.
The pressure trap at the dead-zone inner edge may also promote rocky planetesimal formation \citep{Ueda+2019,Ueda+2021}.
However, recent two-dimensional radiation-hydrodynamic simulations suggest that thermal instability can drive cyclic episodes of MRI activation and deactivation in the inner disk, implying that the location of the MRI activation boundary may not remain steady over time \citep{Cecil&Flock2024, Cecil+2026}.
In such a case, dust accumulation and planetesimal formation at the dead-zone inner edge could be hindered.

\citet{Kato+2025} proposed a novel mechanism, referred to as thermally driven dust accumulation, that causes dust accumulation in a viscously heated disk.
In this mechanism, dust particles can spontaneously accumulate if there is a local enhancement in the dust surface density, even without a pressure bump steadily maintained by a specific gas surface density structure.
In viscously heated disks, the midplane temperature increases with vertical optical depth. Therefore, a local enhancement in the dust abundance raises the disk temperature. This local temperature enhancement, in turn, generates a pressure bump and promotes further dust accumulation. 
\citet{Kato+2025} imposed a local enhancement in the dust surface density as an initial condition, without considering the mechanism that generates it. 

Recurrent MRI activation may provide conditions favorable for thermally driven dust accumulation, although such accumulation has not been observed in previous studies.
In the simulations by \citet{Cecil&Flock2024}, the MRI-active region is unsteady, but a transient gas surface density bump forms at its outer edge, producing a corresponding pressure bump.
Using one-dimensional radiation-hydrodynamic simulations that include dust growth, radial drift, and turbulent diffusion, \citet{Ziampras+2026} confirmed that recurrent MRI activation occurs and showed that the resulting gas bumps can act as dust traps until they diffuse away.
However, thermally driven dust accumulation is not found in their results. 
This may be attributed to the lower gas surface density adopted in their simulations, which is approximately an order of magnitude smaller than that adopted by \citet{Kato+2025}.
As a result, viscous heating is less efficient, and the dust bump does not significantly alter the temperature structure from one dominated by irradiation heating.
If the gas surface density is higher and the temperature structure changes more significantly due to the dust bump, allowing thermally driven dust accumulation to operate, planetesimals may form spontaneously.

In this study, we demonstrate that thermally driven dust accumulation can occur in the thermally unstable inner regions and may thereby enable planetesimal formation. To this end, we simultaneously calculate non-equilibrium thermal evolution and dust evolution while accounting for MRI activation under high-temperature conditions. 
Our simulations show that the MRI is periodically activated in the inner regions, and that planetesimals form repeatedly following this cycle. The activation and deactivation of the MRI constrain the planetesimal-forming region.

This paper is organized as follows.
Section~\ref{sec:method} describes our numerical model. 
Section~\ref{sec:results} presents the results of our numerical simulations and examines how the planetesimal-forming region and the planetesimal formation efficiency depend on model parameters. 
In section~\ref{sec:discussion}, we discuss the implications of our results for planetary system formation and evaluate the validity and limitations of our modeling assumptions. 
Finally, we summarize the main conclusions in section~\ref{sec:summary}.


\section{Method}~\label{sec:method}
\begin{figure*}
    \begin{center}
     \includegraphics[width = 170mm]{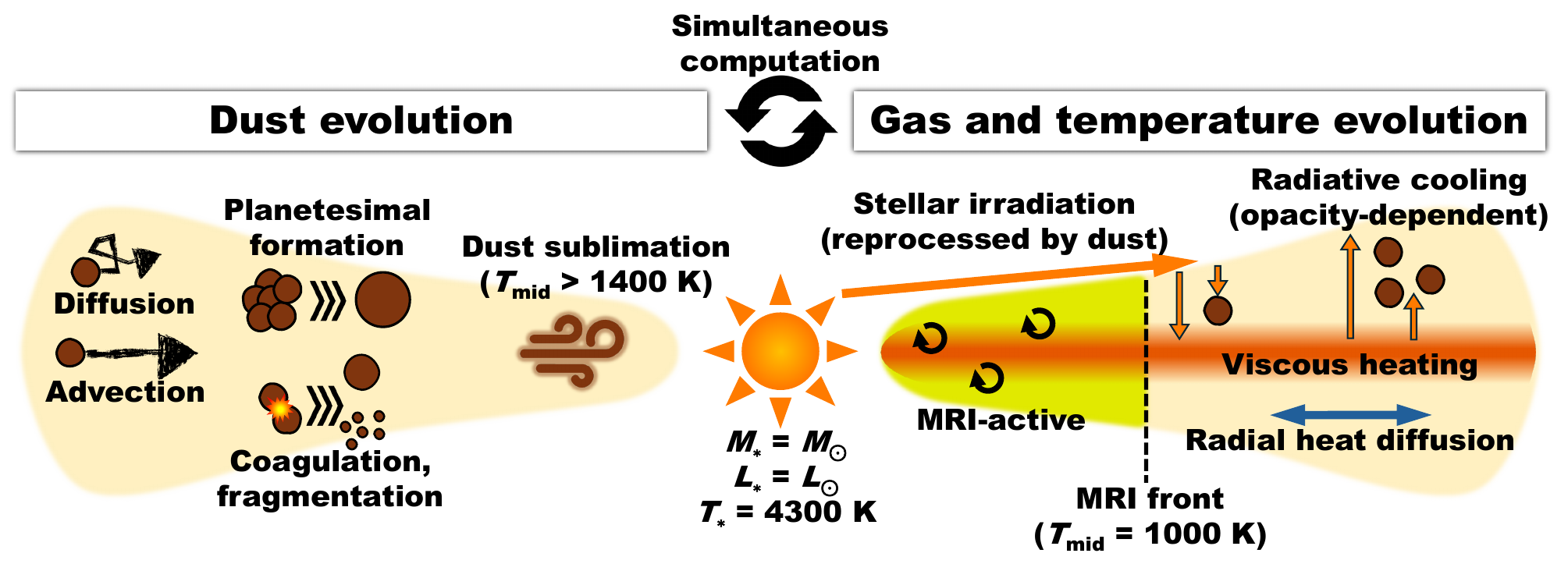}
    \end{center}
    \caption{Schematic illustration of our model to investigate planetesimal formation in thermally unstable protoplanetary disks through the coevolution of gas, dust, and temperature.
%
    }
    \label{fig:model_schematic}
\end{figure*}

This section describes our model for investigating the coevolution of dust, gas, and temperature structures, along with planetesimal formation. 
The model employed in this study is a modified version of that presented by \citet{Kato+2025}. 
We extend the previous work by incorporating physical processes that were not considered in our earlier model.
The key modifications are summarized below. 
For dust evolution, we add the aerodynamic feedback from dust to gas, dust sublimation, and planetesimal formation. In regard to gas evolution, we account for activation of the MRI in high-temperature regions.
In terms of temperature evolution, we incorporate stellar irradiation. 
A detailed description of the baseline model is given by \citet{Kato+2025}.
Figure~\ref{fig:model_schematic} shows a schematic illustration of our model.
\subsection{Dust and gas density evolution}
\label{sec:density_evo}
We calculate the evolution of the gas and dust surface densities, $\Sigmag$ and $\Sigmad$, in one dimension as a function of radial distance $r$ from the central star by solving the continuity equations,
\begin{equation}
    \frac{\del \Sigmag}{\del t}
    +
    \frac{1}{r} \frac{\del}{\del r}\left(r \vgr \Sigmag \right)
    = 0,
    \label{eq:Sigmag_evo}
\end{equation}
and
\begin{equation}
    \frac{\del \Sigmad}{\del t}
    +
    \frac{1}{r}\frac{\del}{\del r}
    \left[
        r \vdr \Sigmad - 
        D_{\rm d} r\Sigmag \frac{\del}{\del r}\left(
        \frac{\Sigmad}{\Sigmag} 
        \right)
    \right]
    = 0,    
    \label{eq:Sigmad_evo}
\end{equation}
where $\vgr$, $\vdr$, and $D_{\rm d}$ are the gas radial velocity, dust radial velocity, and dust diffusion coefficient, respectively.
Accounting for the aerodynamic backreaction from dust to gas, $\vgr$ and $\vdr$ are given by \citep{Kretke+2009,Kanagawa+2017}
\begin{equation}
    \vgr = 
    \left[
    1-\frac{(1+Z)Z}{\St^2+(1+Z)^2}
    \right]v_{\rm vis}
    +\frac{2\St Z}{\St^2+(1+Z)^2}\Delta v_{{\rm g},\phi}
    \label{eq:vgr}
\end{equation}
and 
\begin{equation}
    \vdr = 
    \frac{1+Z}{\St^2 +(1+Z)^2}v_{\rm vis}
    +\frac{2 \St}{\St^2 + (1+Z)^2}\Delta v_{{\rm g},\phi},
    \label{eq:vdr}
\end{equation}
respectively.
Here, $\St$, $Z$, $v_{\rm vis}$, and $\Delta v_{{\rm g},\phi}$ are the Stokes number, the dust-to-gas surface density ratio, the radial gas velocity driven by viscous diffusion, and the deviation of the gas azimuthal velocity from the Keplerian velocity $v_{\rm K}$, respectively.
{\bf Equations~\eqref{eq:vgr} and \eqref{eq:vdr} assume the terminal-velocity approximation, which is typically valid when ${\rm St} \ll 1$. }
One can express $v_{\rm vis}$ and $\Delta v_{{\rm g},\phi}$ as 
\begin{equation}
    v_{\rm vis} = 
    -\frac{3 \nu_{\rm t}}{r} \frac{\del \ln({r^{1/2} \nu_{\rm t} \Sigmag})}{\del \ln{r}},   \label{eq:vvis}
\end{equation}
with $\nu_{\rm t}$ being the gas turbulent viscosity coefficient, and 
\begin{equation}
    \Delta v_{{\rm g},\phi} = \frac{1}{2}\frac{\cs^2}{\vK}\frac{\del \ln{\pmid}}{\del \ln{r}}
    \label{eq:etavK}
\end{equation}
with $\pmid$ being the midplane gas pressure, respectively.
The gas pressure is expressed as $p=c_{\rm s}^2\rho_{\rm g}$, where $c_{\rm s}$ and $\rho_{\rm g}$ are the sound speed and gas density, respectively.
For the dust diffusion coefficient, we use \citep{Youdin&Lithwick2007}
\begin{equation}
    D_{\rm d} = \frac{D_{\rm g}}{1+\St^2},
\end{equation}
with $D_{\rm g}$ being the gas diffusion coefficient. Following \citet{Okuzumi&Hirose2011}, we express $D_{\rm g}$ by using $\nu_{\rm t}$, as $D_{\rm g}=0.3\nu_{\rm t}$.

We approximate the particle Stokes number using its midplane value. To incorporate both the Epstein and Stokes regimes, we apply \citep[e.g.,][]{Birnstiel+2010}
\begin{equation}
    \St = \frac{\pi}{2}\frac{\rho_{\rm {int}} a}{\Sigmag}\max\left(1, \frac{4a}{9\lambda_{\rm mfp}}\right),
    \label{eq:St}
\end{equation}
where $\rho_{\rm int}$ and $a$ are the dust particle's internal density and radius, respectively, and $\lambda_{\rm mfp}$ is the mean free path of the gas molecules, written as $\lambda_{\rm mfp}=m_{\rm g}/(\sigma_{\rm mol}\rho_{\rm g,mid})$, with $m_{\rm g}=3.82\times10^{-24}~\rm g$, $\sigma_{\rm mol}=2.0\times10^{-15}~\rm cm^2$, and $\rho_{\rm g,mid}$ being the mean molecular mass, the molecular collision cross section, and midplane gas density, respectively. 

Our model is essentially one-dimensional in the radial direction and does not require a detailed model for the vertical disk structure. For this reason, we assume that the disk is vertically isothermal and in hydrostatic equilibrium. These two approximations yield the following analytical expression for the vertical distribution of the gas density: 
\begin{equation}
    \rho_{\rm g}(r,z) = 
    \rho_{\rm g,mid}(r) 
    \exp{\left[-\frac{z^2}{2 h_{\rm g}(r)^2}\right]},
    \label{eq:rho_g_z}
\end{equation}
where $z$, $\rho_{\rm g,mid}$, and $h_{\rm g}$ are the height above the midplane, the midplane gas density, and the gas scale height for a disk that is vertically isothermal at the midplane temperature $\Tmid$, respectively, and $\rho_{\rm g,mid}$ is written as
\begin{equation}
    \rho_{\rm g,mid}(r) = \frac{\Sigmag(r)}{\sqrt{2 \pi}h_{\rm g}(r)}.
\end{equation}
The gas scale height is related to the isothermal sound speed $\cs$ at the midplane and Keplerian frequency $\Omega_{\rm K}$ as $h_{\rm g} = \cs/\Omega_{\rm K}$. The isothermal sound speed can be written as $\cs=\sqrt{k_{\rm B}\Tmid/m_{\rm g}}$, where $k_{\rm B}$ is the Boltzmann constant.

For gas turbulent viscosity, we employ the $\alpha$-prescription \citep{Shakura+1973}
\begin{equation}
    \nu_{\rm t} = \alpha c_{\rm s} h_{\rm g},
    \label{eq:nu}
\end{equation}
where $\alpha$ is the dimensionless turbulent viscosity. 
In the model of \citet{Kato+2025}, the radial profile of $\nu_{\rm t}$ is fixed to its initial value. 
In contrast, we account for the evolution of $\nu_{\rm t}$ in response to the temperature evolution. Since $c_{\rm s}$ and $h_{\rm g}$ depend on temperature, $\nu_{\rm t}$ also depends on the disk temperature.  
Furthermore, to model MRI activation, we assume that the local turbulent strength $\alpha_{\rm local}$ depends on the local temperature.
In reality, turbulence generated in MRI-active regions is not strictly localized; wave motions and turbulent eddies can penetrate into neighboring dead zones \citep[e.g.,][]{Fleming&Stone2003,Okuzumi&Hirose2011}. The locally evaluated $\alpha_{\rm local}(r)$ therefore represents only the intrinsic MRI activity, while the effective $\alpha(r)$ should vary more gradually. To model this effect, we smooth the radial profile of $\alpha$ using
\begin{equation}
    \alpha(r,\Tmid(r)) = \frac{\int^{\infty}_{-\infty}\alpha_{\rm local}(r',\Tmid(r')) \exp\left[{-\frac{(r'-r)^2}{2\Delta r^2}}\right]dr'}{\int^{\infty}_{-\infty}\exp\left[{-\frac{(r'-r)^2}{2\Delta r^2}}\right]dr'},
\end{equation}
where $\Delta r$ is the smoothing length, and we set $\Delta r = 2.5h_{\rm g}$ throughout this paper. 
Following the approach of \citet{Flock+2016}, we express the temperature dependence of $\alpha_{\rm local}$ as
\begin{equation}
    \alpha_{\rm local}(\Tmid) = \frac{\alpha_{\rm MRI}-\alpha_{\rm DZ}}{2}
    \left[1-\tanh\left(\frac{T_{\rm MRI}-T_{\rm mid}}{40~{\rm K}}\right)
    \right]+\alpha_{\rm DZ},
\end{equation}
where $\alpha_{\rm MRI}$, $\alpha_{\rm DZ}$, and $T_{\rm MRI}$ are the $\alpha$ values in the MRI-active region and in the dead zone, and the midplane temperature threshold to activate MRI, respectively. Throughout this paper, $T_{\rm MRI} = 1000~\rm K$ is adopted \citep{Desch+2015}.

\subsection{Vertical dust distribution}
Assuming the balance between the vertical settling and diffusion of dust particles, we express the dust scale height as \citep{1995Icar..114..237D,Youdin&Lithwick2007}
\begin{equation}
    h_{\rm d} = h_{\rm g} \left(1+\frac{\rm St}{\alpha_{\rm diff}}\frac{1+2\rm St}{1+\rm St}\right)^{-1/2},
\end{equation}
where $\alpha_{\rm diff} \equiv D_{\rm g}/(c_{\rm s}h_{\rm g}) = 0.3\alpha$ is the dimensionless turbulent diffusion coefficient. 
We assume that the vertical distribution of the dust density follows a Gaussian profile \citep{Carballido+2006}
\begin{equation}
    \rho_{\rm d}(r,z) = 
    \rho_{\rm d,mid}(r) 
    \exp{\left[-\frac{z^2}{2 h_{\rm d}(r)^2}\right]},
    \label{eq:rhod_z}
\end{equation}
where $\rho_{\rm d,mid}$ is the midplane dust density, given by
\begin{equation}
    \rho_{\rm d,mid}(r) = \frac{\Sigmad(r)}{\sqrt{2 \pi}h_{\rm d}(r)}.
\end{equation}

\subsection{Dust particle-size evolution}
\label{sec:size_evo}
For the dust particle-size evolution, we use the same model adopted by \citet{Kato+2025}, which employs the single-size approach of \citet{Sato+2016}. This approach assumes that the dust mass budget is dominated by particles of a representative mass $m_{\rm p}$, which depends on $t$ and $r$. 
We calculate the evolution of $m_{\rm p}$ by solving
\begin{equation}
    \frac{\del m_{\rm p}}{\del t}
    +
    v_{{\rm d},r}\frac{\del m_{\rm p}}{\del r}
    =
    \epsilon_{\rm grow} \frac{2 \sqrt{\pi} a^2 \Delta v_{\rm pp}}{h_{\rm d}} \Sigma_{\rm d},
\end{equation}
where $\epsilon_{\rm grow}$, $\Delta v_{\rm pp}$, and $h_{\rm d}$ are the sticking efficiency, collision velocity, and scale height of the mass-dominating particles, respectively.
The sticking efficiency $\epsilon_{\rm grow}$ is written as \citep{Okuzumi+Hirose2012}
\begin{equation}
    \epsilon_{\rm grow} = \min \left[ 1, -\frac{\ln{(\Delta v_{\rm pp}/v_{\rm frag}})}{\ln{5}} \right],
\end{equation}
where $v_{\rm frag}$ is the threshold collision velocity above which the colliding particles fragment, i.e.,  $\epsilon_{\rm grow} < 0$.

Our model also incorporates dust sublimation. We assume that almost all dust particles sublimate in the regions where the midplane temperature exceeds $1400~\rm K$, as prescribed by 
\begin{align}
    \Sigmad &= 10^{-10}\Sigmag, \\
    a &= a_0,
\end{align}
where $a_0=0.1~{\mu\rm m}$ is the minimum value of $a$. Here, a tiny amount of dust is retained to ensure numerical stability \citep{Ueda+2019}.

\subsection{Planetesimal formation}
\label{sec:plts_model}
In regions where the midplane dust-to-gas mass ratio $\rho_{\rm d,mid}/\rho_{\rm g,mid}$ exceeds unity, we convert a fraction of dust into planetesimals according to \citep{Drążkowska+2016}
\begin{equation}
    \frac{\del \Sigma_{\rm plts}}{\del t} = \zeta\frac{\Sigma_{\rm d}}{T_{\rm K}},
    \quad
    \frac{\del \Sigma_{\rm d}}{\del t} = -\frac{\del \Sigma_{\rm plts}}{\del t}
    \label{eq:plts}
\end{equation}
where $\Sigma_{\rm plts}$, $T_{\rm K}$, and $\zeta$ are the planetesimal surface density, the orbital period, and a dimensionless constant. The quantity $T_{\rm K}/\zeta$ represents the timescale on which dust is converted into planetesimals in the high $\rho_{\rm d,mid}/\rho_{\rm g,mid}$ regions.
The timescale for the development of strong clumping and the fraction of dust converted into planetesimals via the streaming instability remain under debate \cite[e.g.,][]{Simon+2016, Chen&Lin2020, Li&Youdin2021, Lim+2025}.
\citet{Lim+2025} found that strong clumping emerges after $\sim 10^3$--$10^4$ orbital periods for particles with ${\rm St} \sim 0.01$.
They also showed that only a small fraction of the total dust mass resides in the high-density tail of the particle density distribution that reaches or exceeds the Roche density.
\citet{Simon+2016} reported that, for particles with ${\rm St} = 0.3$, a few tens of percent of solids are converted into planetesimals within a few tens of orbital periods. However, simulations that include smaller particles \citep[e.g.,][]{Lim+2025} or a particle size distribution \citep[e.g.,][]{Matthijsse+2025} suggest that the conversion efficiency may be reduced.
Taking these considerations into account, we adopt $\zeta = 10^{-4}$. 

\subsection{Temperature evolution}
\label{sec:T_evo}
We employ the temperature evolution model of \citet{Kato+2025}, modified to account for stellar irradiation heating. The irradiation flux is computed using the widely used two-layer disk model, in which the disk surface absorbs stellar radiation and reprocesses it as infrared radiation that heats the disk interior \citep{Kusaka+1970,Chiang_1997, Chiang+2001}.
To calculate the evolution of the midplane temperature, we use a vertically integrated energy equation that accounts for stellar irradiation heating, viscous heating, vertical radiative cooling, radial heat diffusion, and radiation from the parent molecular cloud, expressed as \citep[e.g.,][]{Watanabe+1990, Watanabe+2008, Okuzumi+2022, Kato+2025}
    \begin{align}
    \frac{\gamma+1}{2(\gamma-1)}\frac{k_{\rm B}\Sigmag}{m_{\rm g}}\frac{\del \Tmid}{\del t} 
    =&
    2C\left[F_{\rm rep,\downarrow} + \sigma_{\rm SB} (T_{\rm ex}^4-\Tmid^4)\right]\notag\\
    &-
    \frac{1}{r}\frac{\del }{\del r}
    \left(r F_{\rm diff}\right) 
    +
    \frac{9}{4}\Sigmag \nu_{\rm t} \Omega_{\rm K}^2,
    \label{eq:energy_eq}
    \end{align}
where $\gamma = 1.4$ is the adiabatic index, $C$ is a dimensionless factor that corrects for the optical thickness and albedo effects on infrared radiation \citep[equation (22) of][]{Okuzumi+2022}, $F_{\rm rep, \downarrow}$ is the vertical downward flux of the reprocessed radiation from the irradiation surface toward the disk interior, $\sigma_{\rm SB}$ is the Stefan-Boltzmann constant, $T_{\rm ex}=10~{\rm K}$ is the radiation temperature of the parent molecular cloud \citep{Ueda+2021_2}, and $F_{\rm diff}$ is the vertically integrated radial heat diffusion flux \citep{Kato+2025}.
The factor $C$ depends on the disk surface density and on the particle absorption and extinction opacities, both of which depend on the calculated particle size (for details, see subsection 2.3 of \citealt{Kato+2025}). In our simulations, the dust is optically thick to its own radiation, and also the absorption and extinction opacities are comparable. In this case, we approximately have $C \approx 8/(3\chi_{\rm R}\Sigma_{\rm g})$, where $\chi_{\rm R}$ is the Rosseland-mean extinction opacity per unit gas mass. 

The downward reprocessed radiation flux is given by
\begin{equation}
    F_{\rm rep,\downarrow} = f_{\downarrow}\mu_* F_* ,
\end{equation}
where $\mu_* = 0.03$, $F_*$, and $f_{\downarrow} = 1/2$ are the sine of the grazing angle between the starlight and disk surface, the magnitude of the direct stellar irradiation flux, and the fraction of the irradiation flux reprocessed downward, respectively. 
For a central star with luminosity $L_*$, $F_*$ can be written as 
\begin{equation}
    F_* = \frac{L_*}{4\pi[r^2 + z_{\rm s}(r)^2]} \approx \frac{L_*}{4\pi r^2},
\end{equation}
where $z_{\rm s}$ is the height of the disk's irradiated surface at radius $r$.
In reality, $\mu_*$ should vary with the disk geometry and the radial distance. Moreover, if the inner disk is puffed up, it can cast a shadow on the outer regions \citep{Dullemond+2001,Dullemond+2004a,Dullemond+2004b}. 
If the main heat source of the disk were stellar irradiation, the temperature in the shadowed regions would become significantly lower than in other regions, potentially affecting the disk's density structure and the compositions of the gas and dust \citep{Wu&Lithwick2021, Ohno&Ueda2021,Notsu+2022}.
However, in the inner regions of interest, viscous heating is the dominant heat source as long as the disk evolves viscously, and stellar irradiation plays only a minor role. For this reason, assuming a constant $\mu_*$ suffices for our model.

\subsection{Initial conditions}
\label{sec:ini_con}
We assume a steady accretion disk in thermal equilibrium as the initial condition. At the beginning of each simulation run, we set the initial midplane temperature $T_{{\rm mid},0}$ as
\begin{equation}
    T_{{\rm mid},0} = (T_{{\rm mid,vis},0}^{4}+T_{{\rm mid,irr},0}^{4} + T_{\rm ex}^4)^{1/4},
\end{equation}
where $T_{{\rm mid,vis},0}$ and $T_{{\rm mid,irr},0}$ are the midplane temperatures determined by the balance between viscous heating and radiative cooling, and between irradiation heating and radiative cooling, respectively.
We adopt $T_{{\rm mid,vis},0}$ and $T_{{\rm mid,irr},0}$ from the temperature models presented by \citet{Oka+2011} and \citet{Ida+2016}, and \citet{Chiang_1997}, respectively: 
\begin{align}
    T_{{\rm mid,vis},0} = 250
    &\left(\frac{M_*}{M_{\odot}}\right)^{3/10}
    \left(\frac{\dot{M}_{\rm st}}{10^{-8}M_{\odot}~{\rm yr^{-1}}}\right)^{2/5} \notag\\
    &\times\left(\frac{\alpha_{\rm DZ}}{10^{-3}}\right)^{-1/5}
    \left(\frac{r}{1~{\rm au}}\right)^{-9/10}~{\rm K}, 
\end{align}
and 
\begin{equation}
    T_{{\rm mid,irr},0} = 150
    \left(\frac{L_*}{L_{\odot}}\right)^{2/7}
    \left(\frac{M_*}{M_{\odot}}\right)^{-1/7}
    \left(\frac{r}{1~{\rm au}}\right)^{-3/7}~{\rm K},
\end{equation}
where $M_*$ is the mass of the central star, and $\dot{M}_{\rm st}$ is the gas accretion rate in a steady state. The central star is assumed to have a mass of $M_{\odot}$, an effective temperature of 4300~K, and a luminosity of $L_{\odot}$. Then, assuming steady accretion \citep{Lynden-Bell+1974}, we set the initial gas and dust surface densities as
\begin{equation}
    \Sigma_{{\rm g},0} = \frac{\dot{M}_{\rm st}}{3\pi\alpha_{\rm DZ}c_{\rm s}h_{\rm g}}
    \label{eq:inig}
\end{equation}
and
\begin{equation}
    \Sigma_{{\rm d},0} = 0.01\Sigma_{{\rm g},0},
    \label{eq:inid}
\end{equation}
respectively.

We consider three models in which $\alpha_{\rm DZ}$, $\alpha_{\rm MRI}$, and $v_{\rm frag}$ are treated as key parameters, as summarized in table~\ref{table:Models}. 
The values of $\alpha_{\rm DZ}$ and $v_{\rm frag}$ are chosen to satisfy the condition for thermally driven dust accumulation found by \citet{Kato+2025}. Specifically, we keep the ratio $\alpha_{\rm DZ}/v_{\rm frag}$ fixed. In the fragmentation-limited regime, this approximately corresponds to ${\rm St}/\alpha_{\rm DZ}$ decreasing from $\sim 100$ to $\sim 10$ as the temperature increases from $\sim100$ to $\sim1000~{\rm K}$. In this regime, a local enhancement in the dust surface density can heat the disk, produce a pressure bump, and further promote dust accumulation. This process requires an intermediate value of ${\rm St}/\alpha_{\rm DZ}$: if it is too small, turbulent diffusion smooths out the dust bump, whereas if it is too large, the bump drifts inward before the temperature structure can respond. We refer the reader to sections 2.5 and 3.2 of \citet{Kato+2025} for the detailed motivation and justification.

In Models~1 and 3, dust particles enter the Stokes drag regime in the planetesimal-forming region during the MRI-inactive phase. However, because the Stokes number is regulated by the fragmentation limit, the transition between the Epstein and Stokes regimes does not significantly affect dust transport.

The Stokes number in our simulations is typically $\sim10^{-3}$--$10^{-1}$ during the MRI-inactive phase and is much smaller during the MRI-active phase. Since ${\rm St}\lesssim0.1$ throughout our calculations, the terminal velocity approximation adopted in equations~\eqref{eq:vgr} and \eqref{eq:vdr} is expected to remain valid.

Our computational domain ranges from $0.05$ to $50~\rm au$ and is divided into 300 cells on a logarithmic scale. 
We impose boundary conditions such that neither gas nor dust enters the computational domain from the inner or outer boundaries, while both are allowed to leave. 

For each model, we run the simulation over the estimated time required for dust initially located at $50~{\rm au}$ to drift to the inner edge of the disk: $5\times10^4~{\rm yr}$ for Models~1 and 3 and $5\times10^5~{\rm yr}$ for Model~2.

\begin{table}
\caption{Parameter sets adopted in our simulation models.}             
\label{table:Models}      
\centering                          
\begin{tabular}{c c c c}        
\hline                 
Model ID & $\alpha_{\rm DZ}$ & $\alpha_{\rm MRI}$ & $v_{\rm frag}$\\               
\hline                        
    Model 1 & $10^{-3}$ & $10^{-2}$ & $10~{\rm m~s^{-1}}$ \\     
    Model 2 & $10^{-4}$ & $10^{-2}$ & $1~{\rm m~s^{-1}}$ \\   
    Model 3 & $10^{-3}$ & $10^{-1}$ & $10~{\rm m~s^{-1}}$ \\
\hline                                   
\end{tabular}
\end{table}

\section{Results} \label{sec:results}
This section presents our simulation results and describes the processes that occur during planetesimal formation in a thermally unstable disk.
We particularly focus on the relation between thermally unstable regions and planetesimal-forming regions, and on the fraction of inward-drifting dust from the outer regions that is converted into planetesimals.
\subsection{Planetesimal formation through recurrent MRI activation} \label{sec:plts_form}
\begin{figure*}
\begin{center}
    \includegraphics[width = 170mm]{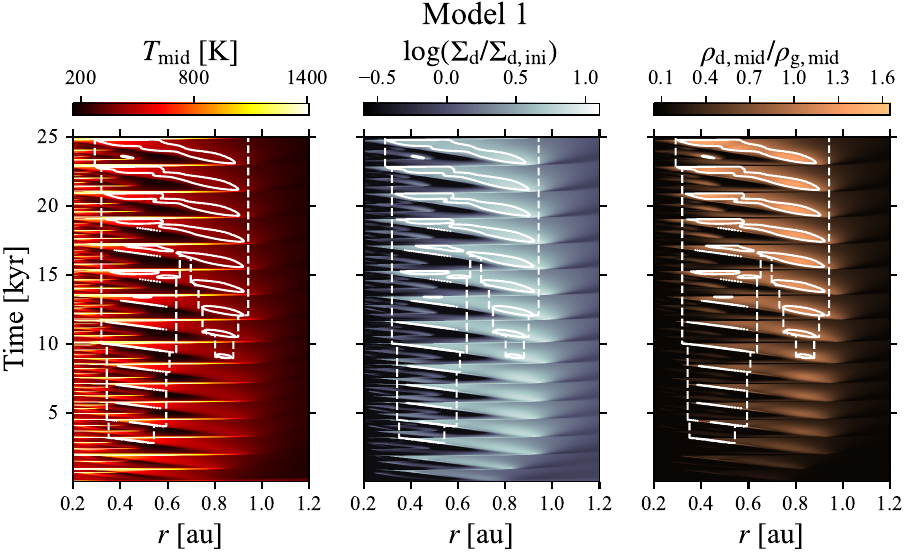}
    \end{center}
    \caption{
    Space-time plots for the midplane temperature $\Tmid$, dust surface density normalized by its initial value $\Sigmad/\Sigma_{\rm d,ini}$, and midplane dust-to-gas mass ratio $\rho_{\rm d,mid}/\rho_{\rm g,mid}$ from Model~1. The regions enclosed by the solid and dashed contour lines represent areas where the midplane dust-to-gas mass ratio exceeds unity, and where planetesimals have formed, respectively.
%
    }
    \label{fig:Model1_T_Sigma_d2g_st}
\end{figure*}

Figure \ref{fig:Model1_T_Sigma_d2g_st} shows the evolution of the midplane temperature $\Tmid$, dust surface density $\Sigmad$, and midplane dust-to-gas mass ratio $\rho_{\rm d,mid}/\rho_{\rm g,mid}$ from Model~1. We observe periodic fluctuations in the temperature and density structures. These fluctuations are caused by thermal instability \citep[e.g.,][]{Armitage2015, Cecil&Flock2024}. In Model 1, planetesimals begin to form at around $t = 2500~\rm yr$. Planetesimal formation also occurs periodically, following the cycle of the thermal instability. 

\begin{figure*}
    \begin{center}
    \includegraphics[width = 170mm]{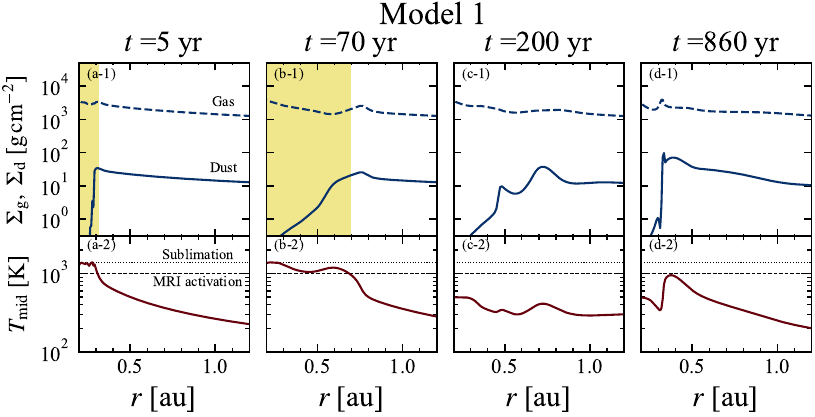}
    \end{center}
    \caption{
    Snapshots of the gas and dust surface densities and midplane temperature from Model~1. The colored regions in the snapshots represent the MRI-active region.
%
    }
    \label{fig:1000}
\end{figure*}

To illustrate a single cycle of the periodic fluctuations, we focus on the first 1000 yr of the simulation in figure~\ref{fig:1000}. First, the MRI becomes active in the hot inner regions ($t=5~{\rm yr}$) and the active region spreads outward ($t=70~{\rm yr}$). 
Panel (b-1) of figure~\ref{fig:1000} shows that a gas surface density bump forms at the outer edge of the active region ($r\approx 0.7~\rm au$).
This bump increases the optical depth just outside the outer edge of the active region, raising the temperature and triggering MRI activation. As a result, the active region expands outward.
At the outer edge of the MRI-active region, a sharp outward drop in turbulent viscosity drives outward angular momentum transport, causing the disk gas to move outward. This can be understood from the fact that when the turbulent viscosity decreases steeply toward larger radii, the derivative term $\del{\ln(r^{1/2}\nu_{\rm t} \Sigmag)}/\del{\ln{r}}$ in equation~\eqref{eq:vvis} becomes negative, resulting in a positive viscous gas velocity. 
This outward gas motion causes gas to pile up at the outer edge, forming a surface density bump.
In the active regions, strong turbulence induces high particle collision velocities, causing particles to fragment into smaller sizes and become more tightly coupled to the gas. Therefore, the dust at the outer edge of the active region is carried outward by the gas, forming a dust bump (panel (b-1)). The MRI-active region stops expanding at the location where the gas and dust surface densities are no longer high enough to sustain the MRI-activation temperature.

Shortly after $t = 70~{\rm yr}$ (panel (b-1) of figure~\ref{fig:1000}), the mass supply of gas to the inner disk decreases, and the inner region can no longer sustain the MRI-active state and returns to the inactive state.
At this point, the dust within the bump at the former outer edge accumulates spontaneously, a phenomenon termed thermally driven dust accumulation \citep{Kato+2025}. 
As the dust bump is more optically thick and cools more slowly than its surroundings, a temperature bump develops there.
This temperature bump, in turn, generates a pressure bump, which further enhances dust accumulation. 
After the MRI becomes inactive ($t = 200~{\rm yr}$, panel (c-1)), the gas accretion rate drops, and the midplane temperature in the inner regions ($r\lesssim 0.7~{\rm au}$) decreases. Meanwhile, the dust bump migrates inward and continues accumulating, causing the temperature to rise again. 
As seen in panels (d-1) and (d-2) at $r \approx 0.35~{\rm au}$, a gas surface density maximum forms just interior to the temperature and dust bumps. This gas bump is produced by the positive radial temperature gradient at the inner edge of the dust bump. This positive temperature gradient enhances the viscosity-driven inward gas velocity, $-v_{\rm vis} \propto \partial \nu_{\rm t}/\partial r \propto \partial T_{\rm mid}/\partial r$ (see equation~\eqref{eq:vvis}), causing a gas pileup. As the dust bump migrates inward, the temperature bump also shifts inward, and the gas surface density bump follows. In this way, the dust bump continues to migrate inward without stalling.
When the temperature in the inner regions reaches 1000~K (see panel (d-2) at $r\approx 0.4~{\rm au}$), the MRI is reactivated, initiating the next cycle. 

\begin{figure*}
    \begin{center}
    \includegraphics[width = 170mm]{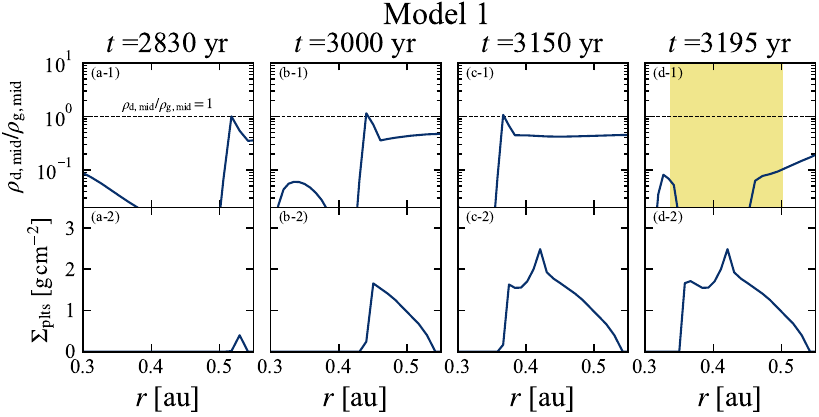}
    \end{center}
    \caption{
    Snapshots of the midplane dust-to-gas mass ratio and planetesimal surface densities from Model~1. The colored regions in the snapshots represent the MRI-active region.
%
    }
    \label{fig:plts}
\end{figure*}

In Figure~\ref{fig:plts}, we show the evolution of the midplane dust-to-gas mass ratio and planetesimal surface density over $t=2830\text{--}3195~{\rm yr}$, focusing on the planetesimal-formation process during repeated MRI cycles.
At $t=2830~{\rm yr}$ (panels (a-1) and (a-2)), the midplane dust-to-gas mass ratio reaches unity and planetesimals begin to form.
Because dust accumulates as the bump migrates inward, planetesimals form from larger to smaller radii. When the MRI is reactivated, planetesimal formation is interrupted and the next cycle begins. In this way, the MRI activation, the outward spreading of the active region, and subsequent inactivation (see also figure~\ref{fig:1000}) regulate where planetesimals form.

\begin{figure}
    \begin{center}
    \includegraphics[width = 75mm]{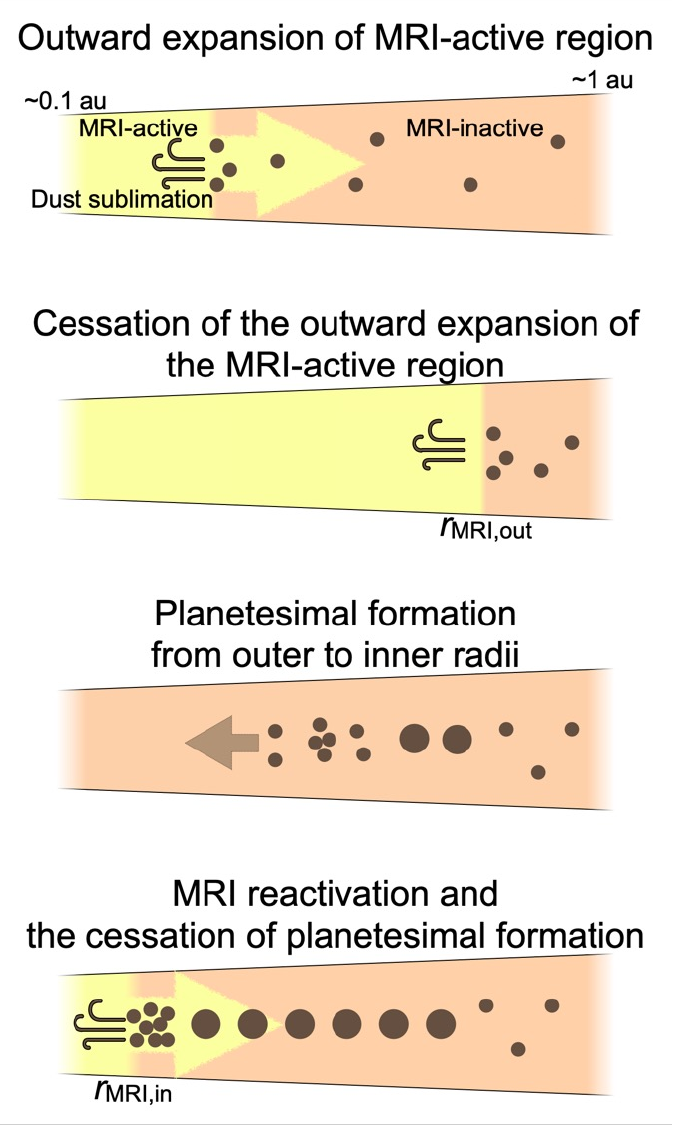}
    \end{center}
    \caption{
    Schematic of planetesimal formation in the thermally unstable inner disk regions.
%
    }
    \label{fig:schematic}
\end{figure}

The planetesimal formation process described above is schematically illustrated in figure~\ref{fig:schematic}. Here, we denote the radius at which the outward propagation of the active region ceases as $r_{\rm MRI,out}$, and the radius at which the MRI is reactivated due to the inward migration of the dust bump as $r_{\rm MRI,in}$.
The outer and inner edges of the planetesimal-forming region correspond to $r_{\rm MRI,out}$ and $r_{\rm MRI,in}$, respectively.

\subsection{Parameter dependence} \label{sec:param_dep}
\begin{figure}
    \begin{center}
    \includegraphics[width = 75mm]{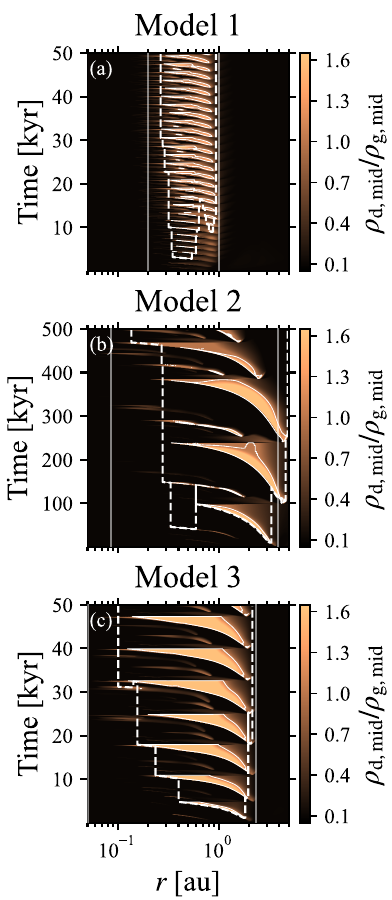}
    \end{center}
    \caption{
    Space-time plots for the midplane dust-to-gas mass ratio from Models~1, 2, and 3, shown in panels (a), (b), and (c), respectively.
    The solid and dashed contour lines are the same as those in figure~\ref{fig:Model1_T_Sigma_d2g_st}.
    The faint gray vertical lines indicate the estimated inner and outer edges of the planetesimal-forming region, corresponding to $r_{\rm MRI,in}$ and $r_{\rm MRI,out}$, respectively.
%
    }
    \label{fig:d2g_all}
\end{figure}

\begin{figure*}
    \begin{center}
    \includegraphics[width = 170mm]{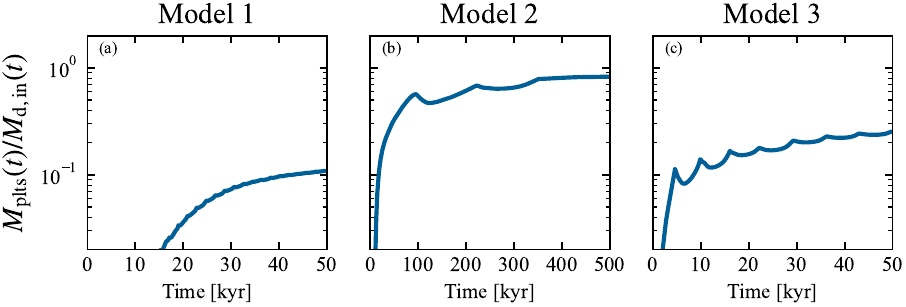}
    \end{center}
    \caption{
    Planetesimal formation efficiency $M_{\rm plts}(t)/M_{\rm d,in}(t)$ as a function of time $t$ from Models~1, 2, and 3, shown in panels (a), (b), and (c), respectively.
    %
    }
    \label{fig:plts_eff_all}
\end{figure*}

Figure~\ref{fig:d2g_all} presents the space-time plots of the midplane dust-to-gas mass ratio from Models~1, 2, and 3. 
We observe that the cycle period and the width of the planetesimal-forming region differ among the models. 
Specifically, Model~1 shows a planetesimal-forming region of $\approx 0.3$--$1$~au with a cycle period of $\approx 2000$~yr, Model~2 shows $\approx 0.1$--$5$~au and $\approx 100000$~yr, and Model~3 shows $\approx 0.1$--$2$~au and $\approx 5000$~yr.
Figure~\ref{fig:plts_eff_all} shows the ratio ${M}_{\rm plts}(t)/{M}_{\rm d,in}(t)$ for the three models, where ${M}_{\rm plts}(t)$ is the total mass of planetesimals formed by time $t$, and ${M}_{\rm d,in}(t)$ is the total mass of dust supplied to the planetesimal-forming region by time $t$. 
We refer to this ratio as the cumulative planetesimal formation efficiency. 
The cumulative planetesimal formation efficiency also differs among the models.
These differences arise from variations in $\alpha_{\rm DZ}$, $\alpha_{\rm MRI}$, and $v_{\rm frag}$. Section~\ref{sec:m1_m2} examines the dependence on $\alpha_{\rm DZ}$ and $v_{\rm frag}$, while section~\ref{sec:m3} examines the dependence on $\alpha_{\rm MRI}$.
\subsubsection{Dependence on $\boldsymbol{\alpha_{\rm DZ}}$ and $\boldsymbol{v_{\rm frag}}$} \label{sec:m1_m2}
As discussed by \citet{Kato+2025}, the efficiency of thermally driven dust accumulation is controlled by the ratio of the viscosity coefficient, corresponding to $\alpha_{\rm DZ}$ in this study, to the dust fragmentation threshold velocity $v_{\rm frag}$. To examine how these parameters affect our simulation results while maintaining the condition for thermally driven dust accumulation, we introduce Model 2, in which both $\alpha_{\rm DZ}$ and $v_{\rm frag}$ are decreased by a factor of 10 from Model 1 while keeping the ratio $\alpha_{\rm DZ}/v_{\rm frag}$ unchanged (see table~\ref{table:Models}). 

Panels (a) and (b) of figure~\ref{fig:d2g_all} present the space-time plots of the midplane dust-to-gas mass ratio from Models~1 and 2, respectively.
We find that, in Model~2, the planetesimal-forming region is wider ($\approx 0.1$--$5$~au) and the cycle period is longer ($\approx 10^5$~yr).
This is because $\alpha_{\rm DZ}$ is smaller in Model~2 than in Model~1.
Because we adopt the same initial gas accretion rate in all models, decreasing $\alpha_{\rm DZ}$ increases the initial gas surface density (see equation~\eqref{eq:inig}) and raises the midplane temperature.
Therefore, the MRI-active region expands to larger radii, leading to a larger $r_{\rm MRI,out}$ and shifting the outer edge of the planetesimal-forming region outward.

By contrast, despite the higher initial gas surface density in Model~2, its $r_{\rm MRI,in}$ is comparable to or even smaller than that in Model~1. This is because the ratio $\alpha_{\rm MRI}/\alpha_{\rm DZ}$ is larger in Model~2. The gas viscous velocity $v_{\rm vis}$ is proportional to $\alpha$ (see equations~\eqref{eq:vvis} and \eqref{eq:nu}). Therefore, before the depletion of the gas surface density in the inner disk becomes significant, the increase in the gas accretion rate associated with MRI activation is proportional to $\alpha_{\rm MRI}/\alpha_{\rm DZ}$. In the inner region where the accretion rate is enhanced, the gas surface density decreases. The degree of this depletion increases with a larger ratio of $\alpha_{\rm MRI}/\alpha_{\rm DZ}$. 
As a result, $r_{\rm MRI,in}$ shifts inward because the gas surface density required for viscous heating to raise the temperature to the MRI activation threshold of $1000~{\rm K}$ is reached at a smaller radius.

{The cycle period is determined by the timescale for dust migration from $r_{\rm MRI,out}$ to $r_{\rm MRI,in}$. 
This dust migration timescale depends on dust drift velocity, which can be approximated as 
\begin{equation}
\vdr \approx 2\St\Delta v_{{\rm g},\phi}.
\label{eq:vdr_approx}
\end{equation}
In our simulations, the maximum dust particle size is regulated by collisional fragmentation induced by turbulence \citep[see equations (26) and (28) of][]{Ormel+2007}, where the Stokes number satisfies $\Delta v_{\rm pp}\approx \sqrt{2.3\alpha \St}c_{\rm s}\approx v_{\rm frag}$ \citep{Birnstiel+2009, Okuzumi&Tazaki2019}. 
Using these approximations, the dust drift speed $|\vdr|$ can be expressed as 
\begin{equation}
    |\vdr| \approx \frac{1}{2.3 v_{\rm K}}
    \left|\frac{\del \ln{p_{\rm mid}}}{\del \ln{r}}\right|
    \frac{v_{\rm frag}^2}{\alpha_{\rm DZ}}.
    \label{eq:vdr_approx_2}
\end{equation}
In Model~2, $v_{\rm frag}^2/\alpha_{\rm DZ}$ is one-tenth as large as in Model~1, and therefore the migration of the dust bump is slower and the cycle period is longer. }
In addition, the wider radial separation between $r_{\rm MRI,out}$ and $r_{\rm MRI,in}$ increases the migration distance of the dust bump and thus further lengthens the cycle period.

A slower migration of the dust bump also affects the efficiency of dust-to-planetesimal conversion.
Panels (a) and (b) of figure~\ref{fig:plts_eff_all} show the cumulative planetesimal formation efficiency, $M_{\rm plts}(t)/M_{\rm d,in}(t)$, from Models~1 and~2, respectively.
We see that the cumulative planetesimal formation efficiency is higher in Model~2. 
Because we adopt the same value of $\zeta$ in all models, the local planetesimal formation timescale $T_{\rm K}/\zeta$ is the same at a given radius across the models (see equation~\eqref{eq:plts}). Therefore, a more slowly migrating dust bump remains longer at each radius, allowing a larger planetesimal mass to form.
This explains why a model with smaller $v_{\rm frag}$ results in more efficient planetesimal formation.
We discuss this in more detail in section~\ref{sec:eff_analytic}.

We note that a secondary dust bump also develops during each MRI activation cycle in Model~2, as shown in panel (b) of figure~\ref{fig:d2g_all}. For example, in the cycle beginning at $t\approx100~{\rm kyr}$, the secondary bump migrates from approximately $2~{\rm au}$ to $0.3~{\rm au}$ over $t\approx100$--$150~{\rm kyr}$. The secondary bump arises from a density perturbation at a smaller radius and can also trigger planetesimal formation.

Nevertheless, the overall MRI cycle period, the location and width of the planetesimal-forming region, and the planetesimal formation efficiency are primarily governed by the main dust bump formed near $r_{\rm MRI,out}$. The cycles at $t\approx100$--$250~{\rm kyr}$ and $250$--$400~{\rm kyr}$ exhibit similar periods and radial extents. We therefore consider the repeated formation and inward migration of the main dust bump to have reached a quasi-periodic state. At later times, however, radial drift depletes dust in the outer disk. Consequently, in the final cycle beginning at $t\approx400~{\rm kyr}$, the reduced dust surface density limits the maximum radial extent of the MRI-active region to approximately $2~{\rm au}$, making this cycle less extended than the previous cycle.

We calculate the Stokes number using equation~\eqref{eq:St}, accounting for both the Epstein and Stokes drag regimes. In Models~1 and 3, which have relatively large values of $v_{\rm frag}$, dust particles grow to sizes of several to approximately $10~{\rm cm}$ in the planetesimal-forming region during the MRI-inactive phase and enter the Stokes drag regime. However, as discussed above, the Stokes number in the region of interest is regulated by the fragmentation limit, and therefore the transition between the drag regimes does not significantly affect dust transport. Although the change in the drag law affects the physical particle size and can thereby modify the opacity, the resulting change in particle size is only by a factor of a few and is therefore unlikely to significantly alter our main results.

\subsubsection{Dependence on $\boldsymbol{\alpha_{\rm MRI}}$} \label{sec:m3}
Panel (c) of figure~\ref{fig:d2g_all} shows the space-time plot of the midplane dust-to-gas mass ratio from Model~3, which adopts $\alpha_{\rm MRI}$ ten times higher than in Model~1. Compared to Model~1 (see panel (a)), Model~3 yields a larger $r_{\rm MRI,out}$. This is because the larger $\alpha_{\rm MRI}$ causes a higher viscous heating rate in the MRI-active state, allowing the disk temperature to reach 1000~K even at lower surface densities.

Model~3 yields a smaller $r_{\rm MRI,in}$ through the same mechanism as Model~2. A larger contrast between $\alpha_{\rm DZ}$ and $\alpha_{\rm MRI}$ produces a larger increase in the accretion rate, depleting gas and dust more significantly. As a result, the radius at which the surface density becomes sufficient for MRI reactivation decreases.

Panel (c) of figure~\ref{fig:plts_eff_all} shows the cumulative planetesimal formation efficiency, $M_{\rm plts}(t)/M_{\rm d,in}(t)$, from Model~3. In Model~3, this efficiency is increased by approximately a factor of two compared to Model~1. This increase arises because a larger contrast between $\alpha_{\rm MRI}$ and $\alpha_{\rm DZ}$ produces more pronounced gas and dust surface density bumps during MRI activation, which enhances thermally driven dust accumulation and leads to a higher dust surface density.

\subsection{Analytic estimates for $\boldsymbol{r_{\rm MRI,out}}$ and $\boldsymbol{r_{\rm MRI,in}}$} \label{sec:analytic}
Here, we analyze in more detail how the inner and outer edges of the planetesimal-forming region are determined.
The outer edge $r_{\rm MRI,out}$ corresponds to the radial distance beyond which the disk can no longer maintain $\Tmid \approx 1000~\mathrm{K}$ in the MRI-active state, whereas the inner edge $r_{\rm MRI,in}$ corresponds to the radial distance at which the temperature reaches 1000 K in the MRI-inactive state under dust-accumulated conditions.
These facts allow us to estimate $r_{\rm MRI,out}$ and $r_{\rm MRI,in}$ analytically.

We begin by relating the midplane temperature $\Tmid$ to the gas surface density $\Sigma_{\rm g}$ under the assumption of local thermal balance. 
In the disk region of interest, viscous heating is the dominant heat source.
The viscous heating and radiative cooling rates per unit area, $Q_{\rm vis}$ and $Q_{\rm cool}$, can be written as
\begin{equation}
    Q_{\rm vis} = \frac{9}{4}\Sigmag\nut\Omega_{\rm K}^2
    \label{eq:Qvis1}
\end{equation}
and
\begin{equation}
    Q_{\rm cool} = 2C\sigma_{\rm SB}T_{\rm mid}^4,
    \label{eq:Qcool1}
\end{equation}
respectively (see the right-hand side of equation~\eqref{eq:energy_eq}).
Substituting equation~\eqref{eq:nu}, $c_{\rm s}=\sqrt{k_{\rm B}\Tmid/m_{\rm g}}$, and $h_{\rm g}=c_{\rm s}/\Omega_{\rm K}$, $Q_{\rm vis}$ can also be written as
\begin{equation}
     Q_{\rm vis} = \frac{9}{4}\alpha\frac{k_{\rm B}\Tmid}{m_{\rm g}}\Omega_{\rm K}\Sigmag.
     \label{eq:Qvis2}
\end{equation} 
We express the cooling rate $Q_{\rm cool}$ based on the formulation of $C$ given by equation~(22) of \citet{Okuzumi+2022}. 
In optically thick regions, where $C \approx 8/(3\chi_{\rm R}\Sigma_{\rm g})$ (see subsection \ref{sec:T_evo}), $Q_{\rm cool}$ is approximated as 
\begin{equation}
    Q_{\rm cool} \approx \frac{16}{3\chi_{\rm R}\Sigma_{\rm g}}\,\sigma_{\rm SB} T_{\rm mid}^4.
    \label{eq:Qcool2}
\end{equation}
By solving the thermal balance equation $Q_{\rm vis} = Q_{\rm cool}$ for $\Sigmag$, we obtain a relation between $\Sigma_{\rm g}$ and $T_{\rm mid}$, 
\begin{align}
    \Sigmag \approx &1400
    \left(\frac{M_*}{M_{\odot}}\right)^{-1/4}
    \left(\frac{\chi_{\rm R}}{1~{\rm cm^2~g^{-1}}}\right)^{-1/2}
    \left(\frac{\alpha}{10^{-2}}\right)^{-1/2} \notag\\
    &\times\left(\frac{\Tmid}{1000~\rm K}\right)^{3/2}
    \left(\frac{r}{1~\rm au}\right)^{3/4}~{\rm g~cm^{-2}}.
    \label{eq:Sigma_eq}
\end{align}
Alternatively, by solving the equation of steady accretion, $\Sigmag=\dot{M}_{\rm st}/(3\pi\nu_{\rm t})$, for $\Tmid$ and eliminating $\Tmid$ from equation~\eqref{eq:Sigma_eq}, we obtain an expression for $\Sigma_{\rm g}$ that involves $\dot{M}_{\rm st}$ instead of $T_{\rm mid}$,
\begin{align}
    \Sigmag \approx & 1600 
    \left(\frac{M_*}{M_{\odot}}\right)^{1/5}
    \left(\frac{\chi_{\rm R}}{0.1~{\rm cm^2~g^{-1}}}\right)^{-1/5}
    \left(\frac{\alpha}{10^{-3}}\right)^{-4/5} \notag\\
    &\times\left(\frac{\dot{M}_{\rm st}}{10^{-8}M_{\odot}~{\rm yr^{-1}}}\right)^{3/5}
    \left(\frac{r}{1~{\rm au}}\right)^{-3/5}
    ~{\rm g~cm^{-2}}.
    \label{eq:Sigma_eq_st}
\end{align}

According to our simulations, the MRI-active zone has a nearly constant temperature of $T_{\rm MRI} \approx 1000~\rm K$, whereas the dead zone is nearly steady with a nearly constant $\dot{M}_{\rm st} \approx 10^{-8}M_{\odot}~\rm yr^{-1}$. Therefore, $r_{\rm MRI,out}$ can be estimated by applying  equations~\eqref{eq:Sigma_eq} and \eqref{eq:Sigma_eq_st} to the active and dead zones ($r \leq r_{\rm MRI,out}$ and $r \geq r_{\rm MRI,out}$), respectively, and equating the two expressions for $\Sigma_{\rm g}$ at $r = r_{\rm MRI,out}$. 
Based on our simulations, we estimate the Rosseland-mean opacities for the MRI-active and dead zones to be $\chi_{\rm R,MRI}\sim1~\rm cm^2~g^{-1}$ and $\chi_{\rm R,DZ}\sim0.1~\rm cm^2~g^{-1}$, with the higher value in the active zone reflecting smaller particle sizes caused by turbulence-induced collisional fragmentation. 
Substituting $\Tmid = 1000~{\rm K}$, $\alpha = \alpha_{\rm MRI}$, and $\chi_{\rm R} = \chi_{\rm R,MRI}$ into equation~\eqref{eq:Sigma_eq}, and $\alpha = \alpha_{\rm DZ}$ and $\chi_{\rm R} = \chi_{\rm R,DZ}$ into equation~\eqref{eq:Sigma_eq_st}, and then equating the two expressions, we have
\begin{align}
    r_{\rm MRI,out} \approx& 1
    \left(\frac{M_*}{M_{\odot}}\right)^{1/3}
    \left(\frac{\dot{M}_{\rm st}}{10^{-8}M_{\odot}~{\rm yr^{-1}}}\right)^{4/9}\notag\\
    &\times\left(\frac{\alpha_{\rm MRI}}{10^{-2}}\right)^{10/27}
    \left(\frac{\alpha_{\rm DZ}}{10^{-3}}\right)^{-16/27}\notag\\
    &\times\left(\frac{\chi_{\rm R,MRI}}{1~{\rm cm^2~g^{-1}}}\right)^{10/27}
    \left(\frac{\chi_{\rm R,DZ}}{0.1~{\rm cm^2~g^{-1}}}\right)^{-4/27}~\rm au.
   \label{eq:pfout}
\end{align}
Our analytical estimate of $r_{\rm MRI,out}$ is consistent with the maximum radial extent of the MRI-active region reported by \citet{Ziampras+2026}. Their analytical estimate and numerical simulations both indicate that the MRI-active region extends outward to approximately $0.9~{\rm au}$. Indeed, evaluating equation~\eqref{eq:pfout} using parameter values similar to those adopted in their study, with $M_*=M_{\odot}$, $\dot{M}_{\rm st}=10^{-9}M_{\odot}~{\rm yr^{-1}}$, $\alpha_{\rm MRI}=0.1$, $\alpha_{\rm DZ}=10^{-3}$, and $\chi_{\rm R,MRI}=7~{\rm cm^2~g^{-1}}$, yields $r_{\rm MRI,out}\approx0.9~{\rm au}$.

During each MRI-active phase, the enhanced gas accretion depletes the gas in the planetesimal-forming region. As the cycle repeats, the gas surface density profile inherited by the subsequent MRI-inactive phase tends to approach the steady-state profile corresponding to the outer-disk mass supply rate, $\dot{M}_{\rm st}=10^{-8}M_{\odot}~{\rm yr^{-1}}$, and the MRI-active viscosity. We therefore adopt this profile as an approximate reference state for estimating $r_{\rm MRI,in}$.

The inner edge of the planetesimal-forming region $r_{\rm MRI,in}$ corresponds to the location where MRI reactivation occurs, i.e., where the midplane temperature reaches 1000~K in the MRI-inactive state due to the opacity increase caused by the inward migration of the dust bump.
Therefore, we estimate $r_{\rm MRI,in}$ by substituting the parameters for the active and dead zones into equations~\eqref{eq:Sigma_eq_st} and \eqref{eq:Sigma_eq}, respectively, and equating the two expressions for $\Sigmag$ at $r=r_{\rm MRI,in}$. 
In addition, at the time of planetesimal formation, the midplane dust-to-gas mass ratio exceeds unity, which increases the opacity. 
We estimate the opacity in the dust-accumulated regions from our simulation results as $\chi_{\rm R} = \chi_{\rm R,DZ,acm} = 10~{\rm cm^2~g^{-1}}$.
Substituting $\alpha = \alpha_{\rm MRI}$ and $\chi_{\rm R} = \chi_{\rm R,MRI}$ into equation~\eqref{eq:Sigma_eq_st}, and $\Tmid = 1000~{\rm K}$, $\alpha = \alpha_{\rm DZ}$ and $\chi_{\rm R} = \chi_{\rm R,DZ,acm}$ into equation~\eqref{eq:Sigma_eq}, and then equating the two expressions, we have
\begin{align}
    r_{\rm MRI,in} \approx& 0.2
    \left(\frac{M_*}{M_{\odot}}\right)^{1/3}
    \left(\frac{\dot{M}_{\rm st}}{10^{-8}M_{\odot}~{\rm yr^{-1}}}\right)^{4/9}\notag\\
    &\times\left(\frac{\alpha_{\rm MRI}}{10^{-2}}\right)^{-16/27}
    \left(\frac{\alpha_{\rm DZ}}{10^{-3}}\right)^{10/27}\notag\\
    &\times\left(\frac{\chi_{\rm R,MRI}}{1~{\rm cm^2~g^{-1}}}\right)^{-4/27}
    \left(\frac{\chi_{\rm R,DZ,acm}}{10~{\rm cm^2~g^{-1}}}\right)^{10/27}~\rm au.
   \label{eq:pfin}
\end{align}
In practice, however, the inner disk does not necessarily reach this reference profile before the subsequent MRI activation. If the gas surface density during the MRI-inactive phase remains higher than the reference profile, viscous heating raises the midplane temperature to $1000~{\rm K}$ at a larger radius. We therefore interpret the estimate given by equation~\eqref{eq:pfin} as an approximate lower limit on $r_{\rm MRI,in}$.

The estimates obtained above yield $(r_{\rm MRI,out},\,r_{\rm MRI,in}) = (1,\,0.2)$~au, $(4,\,0.09)$~au, and $(2,\,0.05)$~au for Models~1, 2, and 3, respectively, which agree with the simulation results (see figure~\ref{fig:d2g_all}) to within a factor of 2.

\subsection{Analytic estimates for planetesimal formation efficiency}
\label{sec:eff_analytic}
In this subsection, we analytically estimate the planetesimal formation efficiency per MRI cycle.
In our simulations, dust bumps induced by recurrent MRI activation form planetesimals while migrating inward from $r_{\rm MRI,out}$ to $r_{\rm MRI,in}$.
Motivated by this behavior, we first estimate the mass of planetesimals formed during a single MRI activation event.
We then estimate the total mass of dust that flows into the planetesimal-forming region during one cycle, with period $t_{\rm cycle}$, and evaluate the planetesimal formation efficiency per cycle.

Taking the width of the dust bump to be $\Delta r_{\rm b}$, the time required for the bump to pass a given radius is $\Delta t = \Delta r_{\rm b}/|\vdr|$.
Our simulations yield $\Delta r_{\rm b} \sim 0.1~{\rm au}$. 
Using these assumptions together with equation~\eqref{eq:plts}, the mass of planetesimals formed per unit radial distance is given by
\begin{equation}
    2\pi r \frac{\del \Sigma_{\rm plts}}{\del t} {\Delta t} = 2\pi r\zeta\frac{\Sigmad}{T_{\rm K}}\frac{\Delta r_{\rm b}}{|\vdr|}.
    \label{eq:plts_form_per_rad}
\end{equation}
In our simulations, $\Sigmag$ and $\Sigmad$ in the planetesimal-forming region deviate from a power-law profile and instead exhibit an approximately flat radial distribution.
We therefore approximate them as radially constant and adopt $|\del \ln{p_{\rm mid}}/\del \ln{r}|\approx2$ for $|\vdr|$.
Furthermore, to account for dust accumulation in the dust bump, we assume $\rho_{\rm d,mid}/\rho_{\rm g,mid} = (\Sigmad/\Sigmag)/(h_{\rm d}/h_{\rm g}) = 1$ and $(h_{\rm d}/h_{\rm g})\approx \sqrt{\alpha_{\rm DZ}/\St}\approx0.1$, which implies $\Sigma_{\rm d}/\Sigma_{\rm g} \approx 0.1$. 
We then estimate $\Sigma_{\rm d}$ at $r_{\rm MRI,out} \sim 1~{\rm au}$ as
\begin{align}
    \Sigmad \approx &160\left(\frac{M_*}{M_{\odot}}\right)^{1/5}
    \left(\frac{\dot{M}_{\rm st}}{10^{-8}M_{\odot}~{\rm yr^{-1}}}\right)^{3/5}\notag\\
    &\times\left(\frac{\alpha_{\rm DZ}}{10^{-3}}\right)^{-4/5}~{\rm g~cm^{-2}}.
    \label{eq:Sigmad_for_pltseff}
\end{align}
Substituting equations~\eqref{eq:vdr_approx_2},~\eqref{eq:Sigmad_for_pltseff}, and $\zeta = 10^{-4}$ into equation~\eqref{eq:plts_form_per_rad} and integrating from $r_{\rm MRI,in} \sim 0.1~{\rm au}$ to $r_{\rm MRI,out} \sim 1~{\rm au}$, we obtain the mass of planetesimals formed during a single cycle as
\begin{align}
    M_{\rm plts,cycle} \approx &1M_{\oplus}
    \left(\frac{M_*}{M_{\odot}}\right)^{6/5}
    \left(\frac{\dot{M}_{\rm st}}{10^{-8}M_{\odot}~{\rm yr^{-1}}}\right)^{3/5}\notag\\
    &\times\left(\frac{\Delta r_{\rm b}}{0.1~{\rm au}}\right)
    \left(\frac{v_{\rm frag}}{10~{\rm m~s^{-1}}}\right)^{-2}
    \left(\frac{\alpha_{\rm DZ}}{10^{-3}}\right)^{1/5}.
    \label{eq:Mpltscycle}
\end{align}
The mass of planetesimals formed in a single cycle, $M_{\rm plts,cycle}$, primarily depends on $v_{\rm frag}$, with smaller $v_{\rm frag}$ yielding a larger mass of planetesimals (equation~\eqref{eq:Mpltscycle}).
As discussed in section~\ref{sec:m1_m2}, this is because a smaller $v_{\rm frag}$  yields slower migration of the dust bump. 

We note that the scaling in equation~\eqref{eq:Mpltscycle} applies only within the parameter regime considered in this study. As discussed in section~\ref{sec:ini_con}, we choose all models to satisfy the condition for thermally driven dust accumulation shown in \citet{Kato+2025}. If we decrease only $v_{\rm frag}$ while keeping $\alpha_{\rm DZ}$ fixed, the Stokes number decreases and the dust couples to the gas more strongly. In that case, the dust cannot accumulate spontaneously.

In contrast to the stronger dependence on $v_{\rm frag}$, the dependence on $\alpha_{\rm DZ}$ is relatively weak, scaling as $\alpha_{\rm DZ}^{1/5}$.
A smaller $\alpha_{\rm DZ}$ reduces the turbulence-driven collision velocity, increases the Stokes number, and accelerates the migration of the dust bump, which tends to reduce the mass of planetesimals formed at a given location.
At the same time, however, a smaller $\alpha_{\rm DZ}$ also increases the initial dust surface density (see equations~\eqref{eq:inig} and~\eqref{eq:inid}), and these effects largely cancel out, resulting in a weak effective dependence on $\alpha_{\rm DZ}$. 

We then estimate the total mass of dust that flows into the planetesimal-forming region over one cycle, which we denote by $M_{\rm d,in,cycle}$.
To this end, we evaluate the dust mass flux at $r_{\rm MRI,out} \sim 1~{\rm au}$, denoted by $\dot{M}_{\rm d,in}$, and $t_{\rm cycle}$. 
The dust mass flux is given by
\begin{equation}
\dot{M}_{\rm d,in} = 2\pi r |{\vdr}| \Sigmad \approx 2\pi r |2{\rm St}\Delta v_{{\rm g},\phi}| \Sigmad.
\label{eq:Mdot_d_in}
\end{equation}
Approximating the disk outside $r_{\rm MRI,out}$ as a steady accretion disk and assuming a steady-state viscously heated disk at 1~au, we estimate $\dot{M}_{\rm d,in}$ as 
\begin{align}
    \dot{M}_{\rm d,in} 
    &\approx 2\pi r 
    \frac{1}{v_{\rm K}}
    \left|\frac{\partial \ln{p_{\rm mid}}}{\partial \ln{r}}\right| \frac{v_{\rm frag}^2}{2.3 \alpha_{\rm DZ}}
    \frac{10^{-2} \dot{M}_{\rm st} \Omega_{\rm K}}{3\pi \alpha_{\rm DZ} c_{\rm s}^2}\notag\\
    &\approx 0.003 
    \left(\frac{M_*}{M_{\odot}}\right)^{-3/10}
    \left(\frac{\dot{M}_{\rm st}}{10^{-8}M_{\odot}~{\rm yr^{-1}}}\right)^{3/5}\notag\\
    &\quad\times\left(\frac{v_{\rm frag}}{10~\rm m~s^{-1}}\right)^{2}
    \left(\frac{\alpha_{\rm DZ}}{10^{-3}}\right)^{-9/5}
    M_{\oplus}~{\rm yr^{-1}},
    \label{eq:Mdot_d_in_2}
\end{align}
where we assume $\Sigmad/\Sigmag\approx0.01$, as the inflow originates from the steady region where no significant dust accumulation occurs.
This surface density ratio differs from that assumed in equation~\eqref{eq:Sigmad_for_pltseff}, which adopts $\Sigmad/\Sigmag\approx0.1$ to account for dust accumulation within the dust bump.
Although both values are evaluated at $r_{\rm MRI,out}\approx 1~{\rm au}$, the width of the dust bump satisfies $\Delta r_{\rm b}\ll 1~{\rm au}$, which makes this approximation valid.

We estimate $t_{\rm cycle}$ as the time required for the dust bump to migrate from $r_{\rm MRI,out}$ to $r_{\rm MRI,in}$, since this migration time is much longer than the duration of the MRI-active phase.
Under this assumption, $t_{\rm cycle}$ is given by 
\begin{align}
    t_{\rm cycle} &= \int^{r_{\rm MRI,out}}_{r_{\rm MRI,in}}\frac{dr}{|\vdr|}\notag\\
    &\approx 2\times10^3 
    \left(\frac{M_*}{M_{\odot}}\right)^{1/2}
    \left(\frac{v_{\rm frag}}{10~{\rm m~s^{-1}}}\right)^{-2}
    \left(\frac{\alpha_{\rm DZ}}{10^{-3}}\right)~{\rm yr},
    \label{eq:tcycle}
\end{align}
where we use equation~\eqref{eq:vdr_approx_2} with $|\del\ln{p_{\rm mid}/\del\ln{r}}|\approx 2$. 

Multiplying equation~\eqref{eq:Mdot_d_in_2} by equation~\eqref{eq:tcycle} yields an estimate of $M_{\rm d,in,cycle}$ as
\begin{align}
    M_{\rm d,in,cycle} \approx &10 M_{\oplus} 
    \left(\frac{M_*}{M_{\odot}}\right)^{1/5}\notag\\ 
    &\left(\frac{\dot{M}_{\rm st}}{10^{-8}M_{\odot}~{\rm yr^{-1}}}\right)^{3/5}
    \left(\frac{\alpha_{\rm DZ}}{10^{-3}}\right)^{-4/5}.
    \label{eq:M_d_in_cycle}
\end{align}
The dust mass supplied during a single cycle, $M_{\rm d,in,cycle}$, mainly depends on $\alpha_{\rm DZ}$ and increases for smaller $\alpha_{\rm DZ}$. 
As $\alpha_{\rm DZ}$ decreases, $\dot{M}_{\rm d,in}$ increases due to both the higher $\Sigmad$ and the larger $|v_{{\rm d},r}|$ (see equation~\eqref{eq:Mdot_d_in}).
However, the increase in $|v_{{\rm d},r}|$ simultaneously shortens $t_{\rm cycle}$ (see equation~\eqref{eq:tcycle}), so that the net increase in $M_{\rm d,in,cycle}$ is primarily driven by the enhancement of $\Sigmad$. 

Dividing equation~\eqref{eq:Mpltscycle} by equation~\eqref{eq:M_d_in_cycle}, we obtain the planetesimal formation efficiency per cycle as 
\begin{align}
    \frac{M_{\rm plts,cycle}}{M_{\rm d,in,cycle}} \approx 
    &0.1 
    \left(\frac{M_*}{M_{\odot}}\right)
    \left(\frac{\Delta r_{\rm b}}{0.1~{\rm au}}\right)\notag\\
    &\times\left(\frac{v_{\rm frag}}{10~{\rm m~s^{-1}}}\right)^{-2}
    \left(\frac{\alpha_{\rm DZ}}{10^{-3}}\right).
    \label{eq:plts_eff}
\end{align}
For Models 1--3, the above estimate predicts ${M_{\rm plts}}/M_{\rm d,in} \approx 0.1$, $1$, and $0.1$, respectively, which agree with the simulation results (see figure~\ref{fig:plts_eff_all}) to within a factor of 3.

\section{Discussion} \label{sec:discussion}
\subsection{Implications for planet formation}
The results of this study suggest that rocky planetesimals can form via the combination of periodic activation--deactivation of the MRI and thermally driven dust accumulation in the inner disk regions where viscous heating is the dominant heat source. 
This planetesimal formation scenario does not require a steady pressure maximum caused by, e.g., the dead-zone inner edge \citep{Kretke+2009, Ueda+2019, Ueda+2021}.
Planetesimals form during the MRI-inactive phase, and the location of planetesimal formation migrates inward from $r_{\rm MRI,out}$ to $r_{\rm MRI,in}$. During this migration, the temperature at the formation location remains in the range of approximately 200--800~\rm{K}, suggesting that the resulting planetesimals are rocky.

The spatial distribution of planetesimals critically influences the architecture of the subsequently formed planetary system. 
\citet{Hansen2009} showed that the architecture of the solar system terrestrial planets can be reproduced by assuming that the planetesimals are initially confined to a narrow annulus between 0.7--1.0~au (see also e.g., \citealt{Ueda+2021,Izidoro+2022}).
In the context of super-Earth formation, a narrow, ring-like distribution of planetesimals has also been suggested to be favorable \citep[e.g.,][]{Batygin&Morbidelli2023, Shibata&Izidoro2025}.
In our model, planetesimals form within the region between $\sim 0.1$ and $\sim 1~\rm au$, with some variation depending on the disk properties. The inner and outer edges of the planetesimal belt are regulated by the MRI activation and deactivation. 
This distribution is broader than that adopted as the initial condition in the above studies, indicating that subsequent orbital evolution is required to reproduce the orbital architecture of the solar system and super-Earth systems.

\citet{Ogihara+2018b} followed the planetesimal surface density evolution for an initial distribution extending from 0.3--3.0~au, using a disk model that accounts for magnetically driven disk winds. 
Their results showed that outer planetesimals migrate inward due to gas drag, while inner planetesimals can migrate outward if the winds produce a positive radial gradient in the gas surface density in the inner disk.
The positive gradient of the gas surface density also arises in our simulation due to the repeated activation of the MRI in the inner regions (see also \citealt{Cecil&Flock2024,Ziampras+2026}). 
The reduction of the gas surface density in the inner region also prevents planets from undergoing significant Type I migration, which is favorable for a non-resonant configuration commonly seen in close-in super-Earths \citep{Ogihara+2018a}.
While our study does not follow the orbital evolution of planetesimals, the similarity in the gas surface density structure suggests that extending our model to include the dynamical evolution of planetesimals and planets may lead to planetesimal distributions relevant to the formation of terrestrial planets in the solar system, as well as close-in super-Earth systems.
Importantly, our study provides a self-consistent planetesimal distribution based on the coevolution of dust and temperature structures, which can serve as a physically motivated initial condition for investigating planetary system architectures.

We have estimated the efficiency of planetesimal formation in our simulations in section~\ref{sec:eff_analytic}. 
In the fiducial model (Model~1; $\alpha_{\rm DZ} = 10^{-3}$, $\alpha_{\rm MRI} = 10^{-2}$, and $v_{\rm frag} = 10~{\rm m~s^{-1}}$), approximately $1~M_{\oplus}$ of planetesimals are formed per cycle of $\sim 10^3$~yr. 
This timescale is sufficiently short compared to the disk lifetime, suggesting that planetesimal formation in our scenario can produce terrestrial planets, including those in the solar system, even if it operates only over a limited period during disk evolution. 
Furthermore, smaller values of $v_{\rm frag}$ and $\alpha_{\rm DZ}$ (Model~2) enhance the efficiency of planetesimal formation.
When both $v_{\rm frag}$ and $\alpha_{\rm DZ}$ are reduced by an order of magnitude, the mass of planetesimals formed per cycle increases by nearly two orders of magnitude (equation~\eqref{eq:Mpltscycle}), while the period of the MRI activation–deactivation cycle becomes longer by one order of magnitude (equation~\eqref{eq:tcycle}).
As a result, the planetesimal formation efficiency is also enhanced.

\subsection{Impact of stellar mass and gas accretion rate}
\begin{figure}
    \begin{center}
    \includegraphics[width = 75mm]{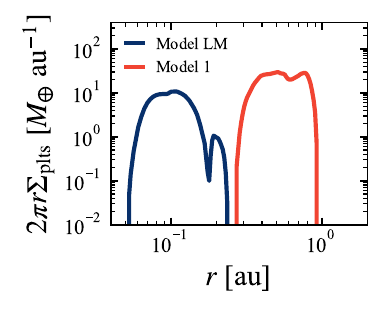}
    \end{center}
    \caption{
    Radial distribution of the planetesimal mass per unit radial distance at $t=50000~\rm yr$ in Models~LM and~1.
%
    }
    \label{fig:lowmass}
\end{figure}
This study focused on dust evolution around a solar-type star with a stellar mass of $M_{\odot}$ and a gas accretion rate of $10^{-8}M_{\odot}~\rm yr^{-1}$. 
However, recent exoplanet surveys have revealed a large population of close-in super-Earths around M-type stars \citep{Mulders+2015b}. 
To examine the effect of the central stellar mass, we reran Model~1 with a stellar mass of $M_* = 0.3M_{\odot}$ (hereafter referred to as Model~LM). The stellar luminosity was set to $L_* = 0.13L_{\odot}$, corresponding to the theoretical estimate for a 3 Myr-old pre-main-sequence star with $M_* = 0.3M_{\odot}$ \citep{Baraffe+2015}.
Using the empirical relationship $\dot{M}_{\rm st}\propto M_*^2$ (\citealt{Manara+2023}), we adopted $\dot{M}_{\rm st}=10^{-9}M_{\odot}~{\rm yr^{-1}}$ for Model LM.
The computational domain was changed to 0.03--30 au in order to capture planetesimal formation in the innermost regions.
Figure~\ref{fig:lowmass} shows the planetesimal mass per unit orbital radius, $2\pi r\Sigma_{\rm plts}$, from Models~LM and~1. 
We see that, in Model~LM, planetesimals form at smaller radii than in Model~1.
This might be linked to the high occurrence rate of close-in super-Earths around low-mass stars \citep{Mulders+2015a}.
With the lower stellar mass and gas accretion rate, the accretion heating is reduced, leading to a lower disk temperature. 
The resulting reduction in the radii of MRI activation and deactivation shifts the planetesimal-forming region inward. 
This behavior can also be understood from the dependencies of equations~\eqref{eq:pfout} and~\eqref{eq:pfin} on $M_*$ and $\dot{M}_{\rm st}$.

\subsection{Impact of dust recondensation}
\begin{figure*}
    \begin{center}
    \includegraphics[width = 160mm]{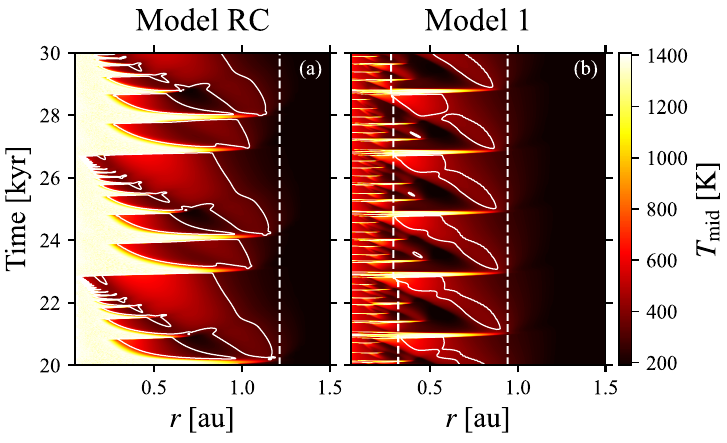}
    \end{center}
    \caption{
    Space-time plots for the midplane temperature $\Tmid$ from Models~RC and 1, shown in panels (a) and (b), respectively.
    The solid and dashed contour lines are the same as those in figure~\ref{fig:Model1_T_Sigma_d2g_st}.
%
    }
    \label{fig:RC}
\end{figure*}
Our simulations include dust sublimation but not dust recondensation. 
In reality, dust recondensation may have a non-negligible impact because 
a large fraction of dust is sublimated in the MRI-active region.
In particular, in Models~1 and~3, where the planetesimal formation efficiency is relatively low, a significant fraction of the inflowing dust undergoes sublimation. 
Multiple scenarios have been proposed for the recondensation of silicate dust, including condensation onto existing particle surfaces \citep[e.g.,][]{Gail&Sedlmayr1999,Ferrarotti&Gail2001} and nucleation from the gas phase \citep[e.g.,][]{Yamamoto&Hasegawa1977,Gail+2013}. However, a unified understanding of which process is dominant has not been established.

Here, we incorporate recondensation into Model~1, hereafter referred to as Model~RC, in order to assess its impact on the results.
For temperatures above 1400 K, we assume complete evaporation of silicate dust and convert the dust surface density into silicate vapor surface density. When the temperature decreases below 1400 K, the vapor is converted back into dust surface density to model recondensation.
Figure~\ref{fig:RC} compares the evolution of the midplane temperature in Models~RC and~1. Including a simplified dust recondensation model slightly modifies the radii at which MRI deactivation and reactivation occur, as well as the cycle period. By accounting for recondensation, dust depletion due to sublimation in the inner region is mitigated, which suppresses the rapid temperature drop. In Model~1, once the MRI-active region reaches $r_{\rm MRI,out}$, the inner region immediately returns to the inactive state. In contrast, in Model~RC, the disk transitions continuously from the outer to the inner regions into the inactive state as the mass supply rate decreases. During this transition, the MRI activation front does not move monotonically inward with time but instead propagates inward while being repeatedly pushed outward. This behavior is essentially the same as the MRI activation “reflares” shown by \citet{Cecil&Flock2024} and \citet{Ziampras+2026}, although the timescale differs.

The planetesimal formation efficiency becomes higher when recondensation is included. This enhancement arises because, during MRI deactivation, dust is replenished in the inner region via recondensation, and the recondensed dust is subsequently incorporated into the inward-migrating dust bump, where it is converted into planetesimals. 
Specifically, the planetesimal formation efficiency reaches approximately 0.5 in Model~RC, approximately five times the value of 0.1 obtained in Model~1, which does not include recondensation.

Dust sublimation and recondensation are important not only for determining the amount of planetesimals formed but also for understanding planetary chemical compositions. 
The difference in sublimation temperatures between silicates and iron may be related to the formation of iron-rich planets, so-called super-Mercuries, which have been discovered in recent years \citep[e.g.,][]{Santerne+18,Johansen&Dorn2022,Mah&Bitsch2023,Tajer+2026}.
In future work, incorporating a more detailed sublimation and recondensation model into our framework may enable a unified treatment of the composition, formation location, and characteristic sizes of terrestrial planets.

\subsection{Regulation of planetesimal formation efficiency}
\begin{figure}
    \begin{center}
    \includegraphics[width = 75mm]{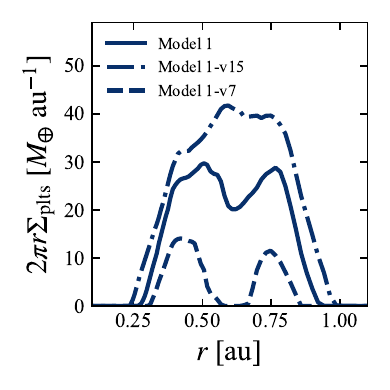}
    \end{center}
    \caption{
    Same as figure~\ref{fig:lowmass}, but from Models~1,~1-v15, and~1-v7. Note that both the horizontal and vertical axes are shown on linear scales.
    %
    }
    \label{fig:vfrag_change}
\end{figure}
In our simulations, approximately 10--80\% of the dust flowing into the planetesimal-forming region is converted into planetesimals. 
Assuming a disk gas mass of $\sim 0.1M_{\odot}$ and a dust-to-gas mass ratio of 0.01, on the order of $100M_{\oplus}$ of dust can potentially be supplied, forming $\sim10$--$80M_{\oplus}$ of planetesimals. This mass is sufficient for the formation of multiple super-Earths.

However, there may be some processes that regulate the dust supply to the inner regions. 
For instance, pressure maxima in the disk outer regions may trap the drifting dust particles, suppressing the dust supply to the inner regions \citep[e.g.,][]{Pinilla+2012,Dullemond+2018}.
Variations in $v_{\rm frag}$ caused by changes in dust composition across the snowline can also affect the dust mass flux into the inner disk.
If icy dust is stickier than silicate dust, the dust supply from the outer regions can be enhanced, since larger particles drift faster.
On the other hand, recent studies have indicated that the adhesion of icy dust is not necessarily higher than that of silicate dust \citep{Gundlach+2018,Musiolik&Wurm2019}.
If icy dust is more fragile than silicate dust, the inward dust mass flux would be reduced, thereby decreasing the amount of planetesimals formed in the inner regions. 

To examine this effect, we perform additional simulations based on the setup of Model~1, in which $v_{\rm frag}$ is modified to $15~\mathrm{m~s^{-1}}$ and $7~\mathrm{m~s^{-1}}$ in regions where $T_{\rm mid}\le 170~\rm{K}$. 
We refer to these models as Models~1-v15 and 1-v7, respectively. 
Figure~\ref{fig:vfrag_change} compares the planetesimal distributions at $t=50000~\rm{yr}$ for these models with that of Model~1. 
The total masses of planetesimals formed in Models~1, 1-v15, and 1-v7 are 13, 21, and 3.5~$M_{\oplus}$, respectively.
In Model~1-v15, the larger value of $v_{\rm frag}$ in the outer cold regions leads to a higher dust supply rate, resulting in a greater total mass of formed planetesimals. 
In contrast, in Model~1-v7, the reduced dust supply from the outer regions yields a smaller planetesimal mass. 

In Models~1 and 1-v7, the planetesimal distribution exhibits a bimodal structure. The outer peak arises because, when the MRI is deactivated, a dust bump forms near $r_{\rm MRI,out}$ and planetesimal formation becomes efficient there. As the bump gradually diffuses and the dust density declines, planetesimal formation becomes less efficient, causing a reduced planetesimal mass interior to the outer peak. Meanwhile, as dust continues to migrate inward while undergoing self-accumulation, the dust density increases again at smaller radii, giving rise to the second, inner peak.
In contrast, in Model~1-v15, the higher dust supply rate from the outer disk produces a more massive dust bump near $r_{\rm MRI,out}$. Consequently, planetesimal formation proceeds continuously from the outer to inner regions, resulting in a singly peaked planetesimal distribution.

\subsection{Survival of formed planetesimals} \label{sec:survival}
In our simulations, we do not follow the orbital or size evolution of the formed planetesimals. Because these planetesimals may be repeatedly exposed to high temperatures of $1400~\rm K$ during subsequent MRI-active phases, they could be eroded or destroyed by sublimation. For the planetesimals to survive subsequent MRI activation events, they must either be transported beyond the thermally unstable region before the next MRI-active phase or have sublimation timescales longer than the duration of the high-temperature phase.

However, it is unlikely that the planetesimals formed by this mechanism migrate beyond the thermally unstable region before subsequent MRI activation. Within the time span of our simulations, the radial pressure gradient in the planetesimal-forming region remains negative when averaged over time. Therefore, outward migration during the planetesimal formation process is not expected. Although planetesimal--planetesimal scattering may cause some radial diffusion, efficient scattering-driven redistribution is expected to occur mainly after the formation of protoplanets \citep[e.g.,][]{Kambara&Kokubo2025}. We therefore do not expect newly formed planetesimals to be scattered out of the thermally unstable region before subsequent MRI cycles.

Nevertheless, we expect that most of the formed planetesimals survive subsequent MRI-active phases. Following \citet{Rafikov&Garmilla2012}, we estimate the sublimation mass flux per unit surface area of a solid at a given temperature $T$ as 
\begin{equation}
    F_{\rm sub}(T) = \langle f_{\rm a} \rangle K_0 e^{-T_0/T},
\end{equation}
where $\langle f_{\rm a} \rangle$ is the sticking probability of vapor molecules impacting the solid surface, and $K_0$ and $T_0$ are material-dependent sublimation parameters. Using the values of $K_0$ and $T_0$ for olivine provided in table 1 of \citet{Rafikov&Garmilla2012}, we estimate the sublimation timescale for a planetesimal of radius $R_{\rm plts}$ as
\begin{align}
    t_{\rm sub}(T_{\rm mid}) &\equiv \frac{m_{\rm plts}}{4\pi R_{\rm plts}^2 F_{\rm sub}(T_{\rm mid})} \notag\\
    &\approx3\times10^{11}
    \left(\frac{\rho_{\rm int}}{3~\rm g~cm^{-3}}\right)
    \left(\frac{R_{\rm plts}}{10~\rm km}\right) \notag\\
    &\quad \times\left(\frac{\langle f_{\rm a} \rangle}{0.1}\right)^{-1}
    \left(\frac{K_0}{1.6\times 10^9~{\rm g~cm^{-2}~s^{-1}}}\right)^{-1} \notag\\
    &\quad\times\exp\left[\frac{T_0}{T_{\rm mid}}-\frac{68100}{1400}\right]~{\rm yr},
    \label{eq:tsub}
\end{align}
where $m_{\rm plts}=(4/3)\pi R_{\rm plts}^3 \rho_{\rm int}$ is the mass of the planetesimal. Although our planetesimal formation prescription does not predict the size of individual planetesimals, numerical simulations of planetesimal formation via the streaming instability suggest typical radii of several tens to hundreds of kilometers \citep[e.g.,][]{Simon+2016,Schäfer+2017,Johansen&Lyra2026}. We therefore adopt $R_{\rm plts}=10~{\rm km}$ as a conservative choice. The sublimation timescale estimated above is far longer than the duration of MRI activation in each cycle, which is $\sim 10^2$ yr in Models 1 and 3 and $\sim 10^3$ yr in Model 2. Moreover, the cumulative duration of the MRI-active phases accounts for only $\sim 0.01$ of the total simulation time. Therefore, the formed planetesimal population is not expected to be completely lost by sublimation before the next cycle, allowing the total planetesimal mass to increase over successive cycles. 


\subsection{Resolution dependence}
\begin{figure*}
    \begin{center}
    \includegraphics[width = 160mm]{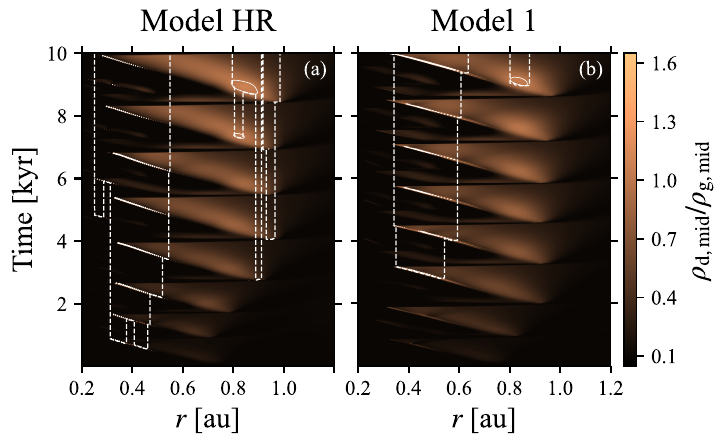}
    \end{center}
    \caption{Space-time plots for the midplane dust-to-gas mass ratio $\rho_{\rm d,mid}/\rho_{\rm g,mid}$ from Models~HR and 1, shown in panels (a) and (b), respectively. The solid and dashed contour lines are the same as those in figure~\ref{fig:Model1_T_Sigma_d2g_st}.
    %
    }
    \label{fig:HR_1}
\end{figure*}
As described in section~\ref{sec:ini_con}, our fiducial simulations use 300 logarithmically spaced radial cells over $0.05$--$50~{\rm au}$. In the inner disk, this corresponds to a ratio of the gas scale height to the radial cell width of $h_{\rm g}/\delta r\sim1$--2. To examine the resolution dependence of our results, we perform a higher-resolution version of Model~1, hereafter referred to as Model~HR. In this model, the radial range $0.1$--$1~{\rm au}$ is divided into 1000 logarithmically spaced cells, yielding $h_{\rm g}/\delta r\sim10$--20. To reduce the computational cost, the computational domain is restricted to $0.1$--$10~{\rm au}$, and the region from $1$ to $10~{\rm au}$ is divided into 100 logarithmically spaced cells. The simulation is evolved for $10~{\rm kyr}$.

Figure~\ref{fig:HR_1} compares the evolution of the midplane dust-to-gas mass ratio in Models~HR and 1. We find no major differences in either the period of MRI activation or the radial extent of the MRI-active region. This agreement indicates that the recurrent MRI activation underlying our planetesimal formation mechanism is qualitatively insensitive to the radial resolution.

The total planetesimal mass formed by $t=10~{\rm kyr}$ is approximately $0.39~M_\oplus$ in Model~HR, compared with $0.13~M_\oplus$ in Model~1. Thus, Model~HR produces approximately three times as much planetesimal mass. This difference likely arises because the reduced numerical diffusion in Model~HR allows a comparable amount of dust to accumulate within a narrower radial region, resulting in a higher local dust density and more efficient planetesimal formation.

These results suggest that the occurrence and cyclic behavior of MRI activation, as well as the operation of the proposed planetesimal formation mechanism, are qualitatively robust against changes in radial resolution. However, quantitative predictions, particularly the peak dust density and the total planetesimal mass, remain sensitive to the numerical resolution. 
Although Model~HR resolves a radial length scale comparable to the gas scale height with 10--20 radial cells, our calculations are vertically averaged and radially one-dimensional. Therefore, caution is needed when interpreting structures on scales smaller than the disk thickness. Resolving such structures more reliably would require two-dimensional calculations that explicitly take the vertical direction into account.

\section{Summary} \label{sec:summary}
We investigated how planetesimals form through the coevolution of dust and disk temperature in the thermally unstable inner regions of protoplanetary disks. To this end, we simultaneously calculated the non-equilibrium thermal evolution, the evolution of the gas and dust surface densities, dust growth, and planetesimal formation, accounting for the MRI activation driven by thermal ionization.

Once the MRI is activated in the inner region, thermal instability drives the outward expansion of the MRI-active zone. 
The expansion halts when it reaches a low-surface-density region where MRI activity can no longer be sustained, after which the inner region returns to an MRI-inactive state.
Such recurrent MRI activity is essentially the same phenomenon as that reported in recent studies, including two-dimensional simulations without dust evolution \citep{Cecil&Flock2024} and one-dimensional simulations with dust evolution \citep{Ziampras+2026}.

In contrast, we demonstrated for the first time that, when the disk returns to an MRI-inactive state, thermally driven dust accumulation is triggered by a dust bump located at the former outer edge of the MRI-active region. 
This is because, in our model, the gas surface density is higher than that adopted by \citet{Ziampras+2026}. 
As a result, viscous heating is more effective, allowing the reduction in radiative cooling efficiency caused by the dust bump to significantly modify the temperature and pressure structures.

The dust bump then migrates inward while accumulating dust, thereby forming planetesimals from larger to smaller radii. 
As the dust bump continues its inward migration, the MRI is reactivated at a smaller radius, which interrupts planetesimal formation.
In this way, the planetesimal-forming region is regulated by the repeated activation and deactivation of the MRI (section~\ref{sec:plts_form}).

The radial extent of the planetesimal-forming region depends on the turbulence strength in both the MRI-dead region, $\alpha_{\rm DZ}$, and the MRI-active region, $\alpha_{\rm MRI}$.
For a given gas accretion rate, a smaller $\alpha_{\rm DZ}$ yields a higher gas surface density, causing the planetesimal-forming region to extend farther outward (section~\ref{sec:param_dep}, equations~\eqref{eq:pfout} and~\eqref{eq:pfin}).
By contrast, a larger $\alpha_{\rm MRI}$ enhances viscous heating in the MRI-active state, allowing the MRI-active region to extend to larger radii and thereby broadening the planetesimal-forming region (equation~\eqref{eq:pfout}).
A larger $\alpha_{\rm MRI}/\alpha_{\rm DZ}$ ratio leads to stronger gas depletion in the inner disk, causing the MRI reactivation radius (i.e., the inner edge of the planetesimal belt) to move inward.
The planetesimal formation efficiency depends on the migration speed of the dust bump, which is determined by $v_{\rm frag}$ and $\alpha_{\rm DZ}$, as well as on the dust surface density, which also depends on $\alpha_{\rm DZ}$ (figure~\ref{fig:plts_eff_all}).
The planetesimal-forming region spans $\sim 0.3$--$1$~au with an efficiency of $\sim 0.1$ in the fiducial model, extends to $\sim 0.1$--$5$~au with a higher efficiency of $\sim 0.8$ for weaker turbulence, and lies within $\sim 0.1$--$2$~au with an efficiency of $\sim 0.2$ for stronger MRI activity.

Our study suggests that, in the thermally unstable inner regions of protoplanetary disks, dust can spontaneously accumulate and form rocky planetesimals even in the absence of a pressure maximum sustained by a steady gas surface density enhancement.
The location and width of the planetesimal-forming region are regulated by repeated MRI activation and deactivation, as well as by the coevolution of dust and temperature.
The physically motivated planetesimal distribution obtained in this study can serve as an initial condition for subsequent planet formation and evolution simulations, providing a pathway toward a framework that consistently tracks the process from dust growth to planet formation.

\section*{Funding}
This work was supported by JSPS KAKENHI Grant Number JP23K25923 and JP26KJ1131, and JST SPRING, Japan Grant Number JPMJSP2180.

\begin{ack}
The authors thank Alexandros Ziampras, Mario Flock, and Michael Cecil for valuable discussions on recurrent MRI activation and the associated disk evolution.
\end{ack}

\bibliographystyle{apj}
\bibliography{paperII}

\end{document}